\documentclass[pdflatex,sn-mathphys-num]{sn-jnl}

\usepackage{graphicx}%
\usepackage{multirow}%
\usepackage{amsmath,amssymb,amsfonts}%
\usepackage{amsthm}%
\usepackage{mathrsfs}%
\usepackage[title]{appendix}%
\usepackage{xcolor}%
\usepackage{textcomp}%
\usepackage{manyfoot}%
\usepackage{booktabs}%
\usepackage{algorithm}%
\usepackage{algorithmicx}%
\usepackage{algpseudocode}%
\usepackage{listings}%
\usepackage[utf8]{inputenc}
\usepackage{subcaption}
\usepackage{float}

\theoremstyle{thmstyleone}%
\theoremstyle{thmstyletwo}%

\theoremstyle{thmstylethree}%

\begin{document}

\title[Article Title]{Localized In-Plane Cavity Optomechanics in MEMS}


\author[]{\fnm{Sasan} \sur{Rahmanian}}\email{s223rahm@uwaterloo.ca}

\affil[]{\orgdiv{Systems Design Engineering Department}, \orgname{University of Waterloo}, \orgaddress{\street{200 University Ave West}, \city{Waterloo}, \postcode{N2L 3G1}, \state{Ontario}, \country{Canada}}}


\abstract{This study demonstrates the realization of localized in-plane optomechanical microcavities embedded within an electrostatic MEMS architecture. The system consists of a curved, clamped-clamped microbeam, fabricated on a silicon-on-insulator (SOI) wafer. A green laser emitted from a Laser Doppler Vibrometer (LDV), is directed perpendicularly onto the device under a vacuum pressure of 7 mTorr, with the beam aligned to fill the gap between the movable microbeam and its adjacent side fixed mirror. This configuration forms localized cavity optomechanical resonators that enable the generation of optomechanical soliton frequency combs through phonon lasing without electrical excitation. The optomechanical resonators\textquotesingle\ dynamics are examined through experiments and numerical simulations. First, the experimental findings unveil that in electrostatic MEMS structures, the two reflective electrodes positioned to form a capacitive gap can inadvertently form localized cavities. These cavities significantly affect optical readouts, as the photodetected signal encodes contributions from both Doppler-shifted electromagnetic waves and light scattered from the intracavity optical field. This dual contributions can distort mechanical response interpretation unless appropriately filtered. Second, experiments show that optical pumping at various positions along the microbeam induces periodic pulse trains with distinct free spectral ranges (FSRs), each corresponding to different mechanical modes. Our results present the generation of solitary optical wavepackets using in-plane localized Fabry-P\'{e}rot microcavities formed within a MEMS device. The results suggest a path toward chip-scale, soliton frequency combs generators featuring frequency spacing on the order of kilohertz, without relying on integrated fiber optics.}




\maketitle

\section{Introduction}\label{sec1}

The generation of stable, phase-coherent pulse trains, manifested as soliton frequency combs (FCs), has become a central focus in nonlinear optics and optomechanics. Recently, Kerr optical frequency combs (OFCs) have leveraged optical solitons to generate stable, well-defined spectral lines, enabling their use as stable frequency baselines with significant advancements in applications such as optical atomic clocks \cite{wu2025vernier, newman2021high, erickson2024atomic}, sensing \cite{xu2021arbitrary, wei2024dual, russell2023tunable}, high-resolution spectroscopy \cite{konnov2023high, ren2023dual}, nonlinear microscopy \cite{chang2024mid, redding2022high}, resonance stabilization \cite{de2023mechanical, long2021electro, stern2020direct}, quantum information processing \cite{zhang2023chip, lu2023frequency}, and telecommunications \cite{oxenlowe2025optical}. These optical solitons emerge in nonlinear media such as integrated waveguides and optical fibers, where a balance between optical dispersion and optical Kerr nonlinearity ensures their temporal and spectral coherence.

Optical frequency combs can be produced through a variety of physical mechanisms, each leveraging distinct principles of light-matter interaction. A widely employed method involves the electro-optic modulation of a continuous-wave (CW) laser source \cite{zhuang2023electro, chen2025microwave, hu2022high}, where an externally applied periodic modulation induces a series of equidistant spectral sidebands around the carrier frequency. Alternatively, comb generation may arise from intrinsic optical nonlinearities, such as four-wave mixing \cite{li2024four, bhardwaj2025generation}, self-phase modulation \cite{roy2024self, shi2025optical}, or cascaded harmonic generation \cite{eliason2024electro, cui2022generation}, within a suitable nonlinear medium. These nonlinear interactions redistribute the optical energy across a broad spectrum, yielding a comb-like structure composed of discrete, equally spaced frequency lines.

Despite the diversity of generation methods, the underlying physics of frequency comb formation can generally be unified under the framework of parametric excitation. In this context, a time-periodic modulation, either externally applied or self-induced through nonlinear back-actions, alters a physical parameter of the system (such as refractive index, cavity length, or gain), resulting in energy transfer across modes and the emergence of dense spectral structures. This modulation-driven mechanism fundamentally describes the appearance of frequency combs across a wide range of platforms, including ring resonators (RRs) \cite{hu2022high, jorgensen2022petabit, cheng2024frequency}, mode-locked lasers (MLL) \cite{trocha2022ultra, riccardi2023short}, and optomechanical cavities \cite{ren2023dual, wan2025optomechanical, kuang2023nonlinear}.

Advances in micro- and nanofabrication have made it possible to realize microscale mechanical structures that interact with light, though their response to direct optical excitation is typically limited due to weak radiation pressure. This challenge is commonly addressed using Fabry-P\'{e}rot optomechanical microcavities, where a movable reflective element is paired with a fixed mirror. The cavity enhances the optical field through constructive interference, increasing the radiation pressure acting on the mechanical structure. This configuration enables effective actuation of the mechanical resonator. These systems enable strong phonon-photon coupling and exhibit autonomous dynamics with cubic nonlinearity.

Recently, optomechanical cavities have been extensively explored for their ability to mediate light-matter interactions. In these systems, the enhanced radiation pressure force arises from the momentum transfer of photons exerted on the movable mirror. This dynamic back-action, where photons interact with mechanical modes, offering prospects for a diverse range of applications: from precision-sensitive measurements to sophisticated quantum manipulation \cite{barzanjeh2022optomechanics} and innovative soliton creation \cite{wang2025retiming}. The formation of optomechanical frequency combs (OMFCs) is contingent upon the optical radiation pressure force exceeding a threshold \cite{miri2018optomechanical}. This condition can be fulfilled, in most fundamental configuration, when the radiation pressure force acting on the mechanical resonator exhibits a purely quadratic dependence on the intracavity optical field, accompanied by a quadratic optomechanical interaction in the optical domain. The optomechanical dynamics can be mapped onto a purely optical framework, wherein the back-action of the radiation pressure force on the optical mode manifests as a non-instantaneous Kerr-like nonlinearity \cite{miri2018optomechanical}. The involvement of mechanical Kerr-like nonlinearities, originating from large structural deformations, also enhances the richness and spectral density of the generated optomechanical FCs for a fixed pump power.

The generation of optical frequency combs using microring resonators \cite{del2007optical, kippenberg2011microresonator} and microtoroids \cite{zhang2021optomechanical, hu2021generation} typically requires ultra-high optical Q-factors, posing significant fabrication and integration challenges. Additionally, the frequency spacing of the comb lines is inversely proportional to the resonators' optical path length (i.e., ring diameter). Consequently, producing equidistant comb teeth with kilohertz-level spacing demands resonator diameters on the order of tens of centimeters, which is impractical for on-chip or compact photonic applications.

In this study, we present a localized cavity optomechanical resonator composed of an in-plane clamped-clamped curved microbeam fabricated on a SOI wafer. A CW laser of fixed wavelength is directed perpendicularly onto the wafer, optically pumping the gap between the suspended microstructure and an adjacent stationary mirror. By enhancing the mechanical Q-factor, achieved through operation under reduced ambient pressure, the intracavity radiation pressure can serve as an optical parametric excitation, effectively actuating the mechanical resonator. This configuration eliminates the need for ultra-high optical Q-factors typically required in conventional microcavity systems. Remarkably, it enables the generation of soliton wavepackets with spectral sidebands spaced in the tens of kilohertz range, using optical pump power of 4 mW.

\section{Results and discussion}\label{sec2}

\begin{figure}[t]
	\centering
	\begin{subfigure}[b]{0.95\textwidth}
		\centering
		\includegraphics[width=\textwidth]{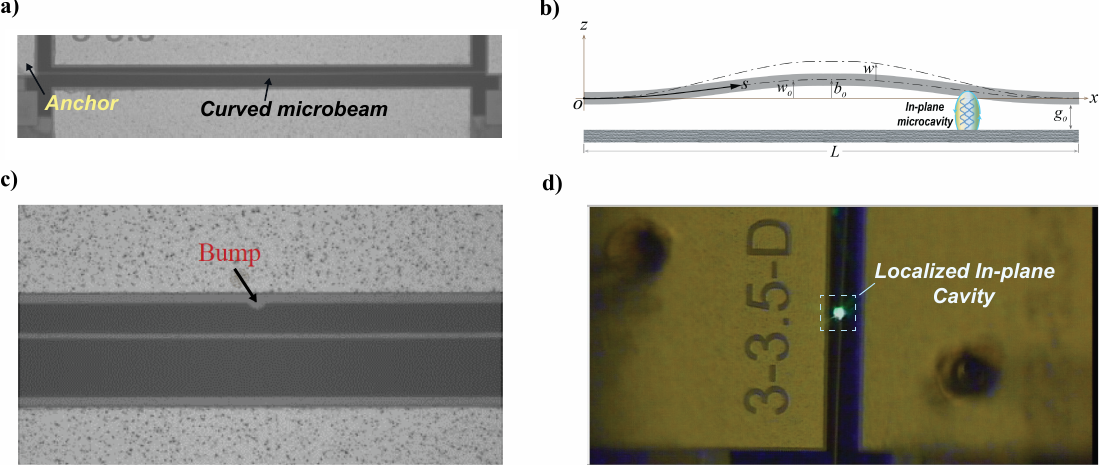}
		\vspace{0.2mm}
	\end{subfigure}
	\caption{Localized in-plane cavity optomechanics. (a) Microscopic image of a clamped-clamped curved microbeam embedded between two actuation electrodes. (b) Schematic of the microbeam resonator coupled to a localized optical cavity. (c) Magnified view of the microbeam midspan obtained using a whitelight profilometer with a 15X objective lens. (d) Optical pumping of the in-plane capacitive gap with a single-wavelength CW green laser at the quarter-length position of the microbeam. The pump laser is directed perpendicular to the silicon substrate, partially illuminating the region between the stationary and movable mirrors, generating an optical radiation pressure force acting on the movable microbeam.}
	\label{fig:1}
\end{figure}

Figure \ref{fig:1}(a) presents a microscopic image of a curved microbeam, anchored at both ends and oriented in-plane, fabricated using the PiezoMUMPs process on a SOI wafer \cite{PiezoMUMPs2014}. It is made from single crystal silicon $<100>$ structural layer with Young\textquotesingle s modulus of $E=128$ GPa, Poisson\textquotesingle s ratio of $\nu=0.22$, and density of $\rho=2230$  kg/m$^3$. The microbeam features the following geometric dimensions: length, $L=1000$ \textmu m, thickness, $h=3$ \textmu m, and width, $b=9$ \textmu m. Figure \ref{fig:1}(b) depicts a schematic of the microbeam. A magnified view showing the midspan vicinity of the curved microstructure is illustrated in Figure \ref{fig:1}(c). This microbeam functions as the movable reflective micromirror in the formation of an in-plane Fabry-P\'{e}rot microcavity. Figure \ref{fig:1}(d) presents the formation of a localized optomechanical cavity when the in-plane gap is optically pumped using a green laser. The gap between the movable mirror and the adjacent stationary mirror, measured at the anchor points, is determined $g_0=10$ \textmu m. The microbeam initial rise at its midspan is defined as $h_0=3$ \textmu m. The microcavity operates within a moderate vacuum, maintained at a pressure of 7 mTorr.

\begin{figure}[t]
	\centering
	\begin{subfigure}[b]{0.97\textwidth}
		\centering
		\includegraphics[width=\textwidth]{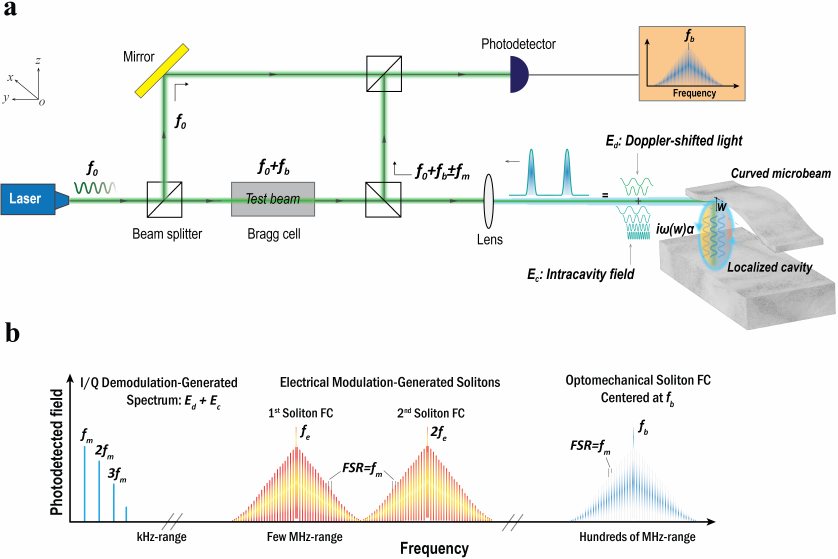}
		\vspace{0.2mm}
	\end{subfigure}
	\caption{(a) Schematic of the experimental setup used to optically pump a MEMS device comprising a clamped–clamped curved microbeam. The microbeam oscillations occur along the z-direction, which lies in-plane with respect to the SOI wafer and is perpendicular to the incident laser beam, propagating along the y-direction. A green, single-wavelength CW laser operates at frequency $f_0=564$ THz and is directed through an acousto-optic Bragg cell, imparting a slight frequency shift of $f_b=621.495$ MHz to generate a test beam at $f_0+f_b$. This beam is focused onto the movable mirror, locally exciting the in-plane microcavity. The total reflected field, comprising both Doppler-shifted light and the intracavity optical field, is collected by the LDV sensor head. A photodetected signal is extracted at the carrier frequency $f_b$ through heterodyne demodulation. (b) Photodetected field spectrum computed by the LDV\textquotesingle s Data Management System (DMS). The spectrum reveals the formation of soliton FCs centered around $f_b$, with equidistant spectral lines locked to a mechanical mode\textquotesingle s natural frequency. Additionally, secondary FCs spanning a few MHz are observed, attributed to electrical modulation occurred in the LDV\textquotesingle s oscilloscope circuitry.}
	\label{fig:2}
\end{figure}

We experimentally demonstrate the formation of localized in-plane optical microcavities within an electrostatic MEMS architecture that operates in the absence of electrical excitation. The microbeam is homogeneous having its neutral axis aligning its middle-plane, featuring a profile $w_0=h_0/2 (1-\cos(2x/L))$ in the rest configuration. The nominal lengths of the localized in-plane cavities vary along different positions of the microbeam span, resulting in distinct fundamental optical resonance frequencies at each location. For instance, the resonance frequency of a cavity formed at a quarter-length of the curved microbeam is determined $f_c=c/(g_0+w_0 (L/4))=25.506$ THz ($c=2.997025\times 10^8$ m/s is the speed of light in air), while that of a cavity located at the microbeam\textquotesingle s midspan is given by $f_c=c/(g_0+w_0 (L/2))=22.2$ THz. The laser pump operates at a fixed frequency of $f_0+f_p$, which is blue-detuned with respect to the resonance frequencies of all in-plane optomechanical cavities formed along the curved microbeam. With this in mind, intracavity photons transfer energy to the mechanical domain through their interaction with phonons, thereby generating coherent phonon populations. This stimulated emission of phonons can be enhanced as the input photon flux increases. When the optical driving power exceeds the intrinsic mechanical dissipation, it gives rise to phase-coherent and self-sustained mechanical oscillations, commonly referred to as phonon lasing. In the present setup, the laser power is held constant at 4 mW, which delivers sufficient optical intensity to satisfy the threshold requirements for initiating optomechanical self-oscillations.
\begin{figure}[t!]
	\centering
	\begin{subfigure}[b]{0.93\textwidth}
		\centering
		\includegraphics[width=\textwidth]{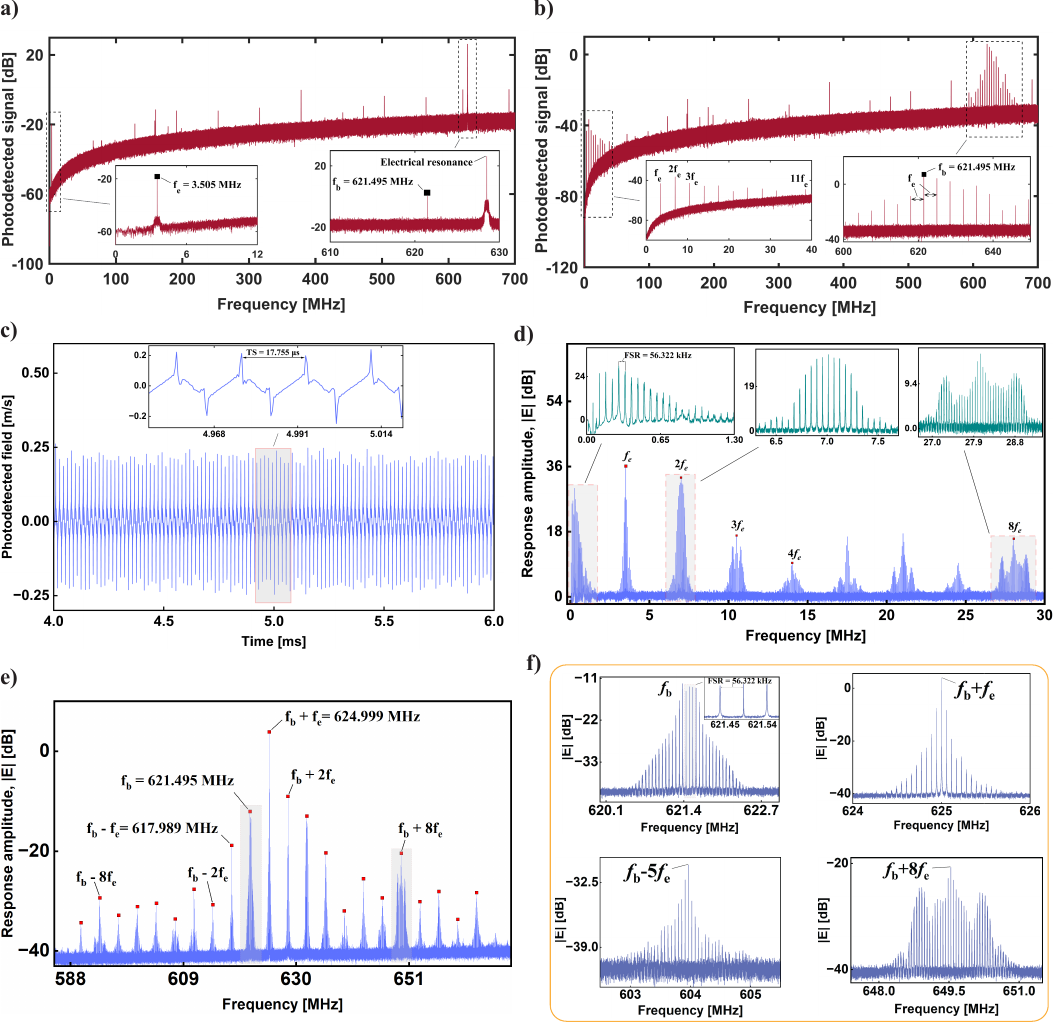}
		\vspace{0.2mm}
	\end{subfigure}
	\caption{Photodetected optical signal. (a) The FFT of the measured signal when the laser is turned off. The harmonic at $f_b=621.495$\ MHz reflects a carrier frequency originating from the photodetector within the LDV\textquotesingle s optical head. (b) Turning on the laser amplifies the component at $f_b$, which is then modulated by the electrical resonance,  $f_e=3.505$\ MHz, generating equidistant FCs around it. The microbeam is pumped at its quarter-length, forming a localized cavity with resonance frequency $f_c=25.506$ THz, under a vacuum pressure of 7 mTorr. (c) Time-domain response exhibiting a pulse train with repetition frequency $f=56.322$ kHz, corresponding to the first anti-symmetric in-plane bending mode of the microbeam. (d) Low-frequency spectral content showing one half of the FCs centered at zero frequency, together with combs centered at $f_e=3.505$ MHz and its higher-order harmonics. (e) High-frequency spectrum showing FCs arsing from electrical modulation of the photodetected signal. (f) Zoomed views of the spectrum at frequencies $f_b$, $f_b+f_e$, $f_b-5f_e$, and $f_b+8f_e$.}
	\label{fig:3}
\end{figure}

Figure \ref{fig:3}(c) presents the measured temporal evolution of the system response acquired when the laser pumps the microcavity at a position corresponding to one-quarter of the curved microbeam\textquotesingle s length, forming a localized cavity with an optical resonance frequency of $f_c=25.506$ THz, under a vacuum pressure of 7 mTorr. The time-domain signal demonstrates a phase-coherent response characterized by a periodic train of pulses, with an inter-pulse interval of TS=17.755 \textmu s. This corresponds to a pulse frequency of $f=56.322$ kHz, aligning with the natural frequency of the microstructure\textquotesingle s second in-plane bending mode. The subplot \uppercase\expandafter{\romannumeral1} in Figure \ref{fig:3}(d) presents the low-frequency range of the optical response, spanning from zero to several hundreds of kilohertz. This spectral content primarily reflects the result of I/Q demodulation applied within the vibrometer\textquotesingle s DMS. Additionally, the localized FCs extending into the megahertz range and centered around the electrical resonance $f_e$ and its higher-order harmonics are attributed to electrical modulation within the oscilloscope\textquotesingle s circuitry.

\begin{figure}[t!]
	\centering
	\begin{subfigure}[b]{0.93\textwidth}
		\centering
		\includegraphics[width=\textwidth]{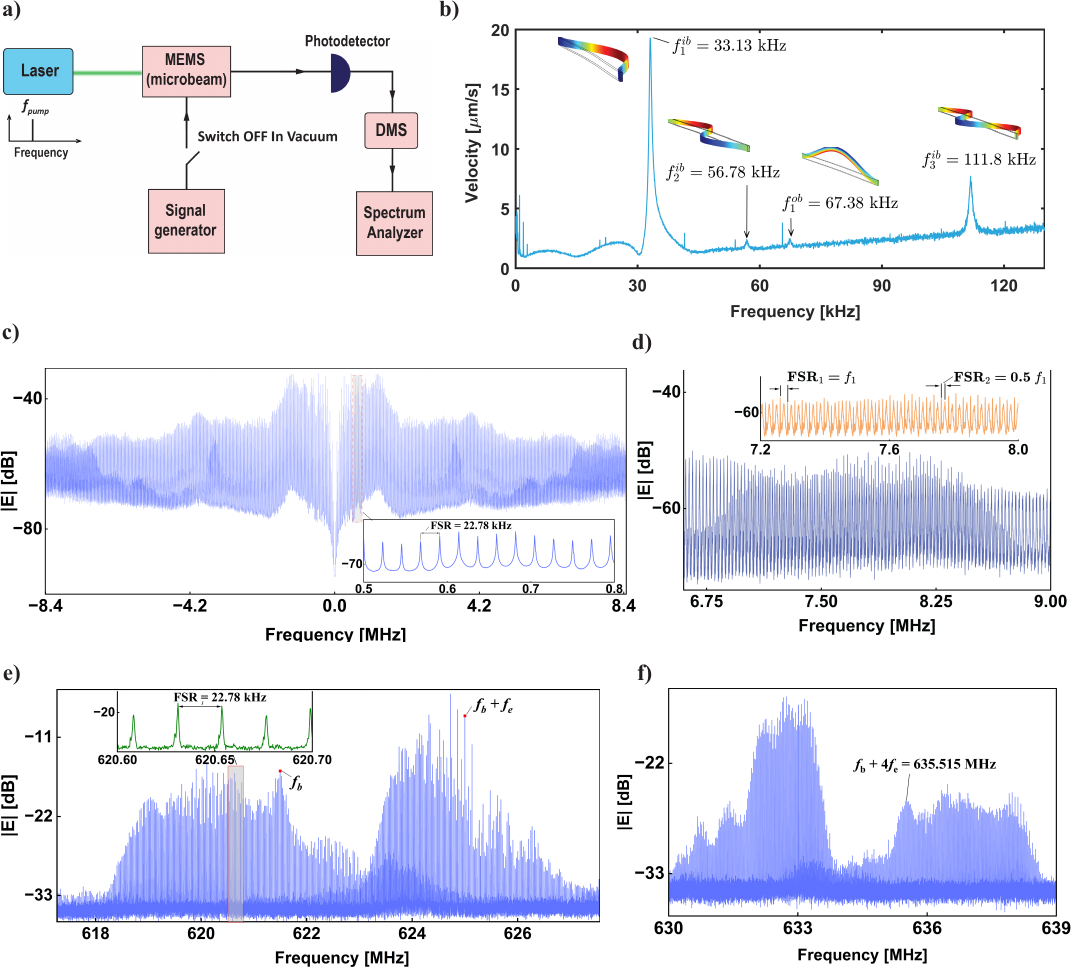}
		\vspace{0.2mm}
	\end{subfigure}
	\caption{(a) Schematic of the experimental setup for optical characterization and microcavity pumping. (b) Mechanical response: FFT of the microbeam\textquotesingle s velocity under an electrostatic pulse (15 V amplitude, 25 Hz frequency, $0.1\%$ duty cycle) measured in air. Absent electrical excitation, optical pumping at three-eighths span generates a localized optomechanical cavity with resonance $f_c = 22.2$ THz under 7 mTorr vacuum.(c) Low-frequency spectrum exhibits densely packed frequency combs (FCs) centered at zero frequency with FSR = 22.78 kHz, corresponding to the tuned first symmetric bending mode. (d) Zoom near twice the electrical resonance ($2f_e = 7.01$ MHz) reveals a dual-comb structure with refined spacing FSR$_2 = 11.39$ kHz. (e) High-frequency spectrum shows soliton FCs centered at $f_b$, merging with electrically modulated FCs at $f_b+f_e$. (f) Detailed view of the comb structure around $f_b+4f_e$.}
	\label{fig:4}
\end{figure}

Figure \ref{fig:3}(e) illustrates the high-frequency portion of the spectrum, spanning from 588 MHz to 570 MHz, which contains optomechanically generated FCs centered at the carrier frequency $f_b$. The measured response reveals the formation of frequency combs induced by electrical modulation, with spectral lines extending up to the eighth harmonic of the electrical resonance frequency and symmetrically distributed about the carrier frequency, $f_b$. Further, in the vicinity of each comb tooth, localized optical wavepackets emerge, exhibiting a fine frequency spacing in the kilohertz range that corresponds to the mechanical oscillation frequency. Magnified spectral views of representative localized comb structures, centered at $f_b$, $f_b+f_e$, $f_b-5f_e$, and $f_b+8f_e$, are depicted in Figure \ref{fig:3}(f). The comb centered at $f_b$ consists of twenty-eight symmetrically distributed, equidistant spectral lines with spacing locked to the natural frequency of the microbeam\textquotesingle s first anti-symmetric in-plane bending mode, $f_2^{ib}=56.322$ kHz, presented in Figure \ref{fig:4}(b). Our experimental observations confirm that laser pumping the microcavity shifts the mechanical resonator static equilibrium, altering the natural frequency of the engaged mechanical mode relative to its unloaded value (see supplementary video).

The intrinsic curvature of the flexible mirror facilitates the formation of localized in-plane optomechanical microcavities with position-dependent resonance frequencies along the microbeam\textquotesingle s span. This spatially varying cavity configuration is particularly advantageous when employing a laser source operating at a fixed wavelength, enabling selective optomechanical interactions without altering the optical source. An enhanced blue-detuning between the resonance frequency of the microcavity and the laser pump frequency is established by directing the laser spot toward the midspan of the curved microbeam. Here, the cavity\textquotesingle s fundamental resonance frequency occurs at $f_c=22.2$ THz. Figure \ref{fig:4}(c) illustrates the low-frequency range of the measured system response $(E)$, revealing a substantial spectral broadening of the optomechanical FCs, extracted through post-photodetection I/Q demodulation. These broadened FCs exhibit significant spectral overlap with those originating from electrical modulation, resulting in the formation of a densely packed optomechanical soliton FCs that extend from DC up to 7.5 MHz. The mechanical oscillation is syntonized to the natural frequency of the microbeam\textquotesingle s first symmetric in-plane bending mode, $f_1^{ib}=22.78$ kHz. Our experimental observation verify that the microstructure undergoes a substantial static deflection with moderate-amplitude oscillations occurring around the newly established equilibrium. Notably, the portion of the laser spot residing in the inter-mirror gap exhibits dynamic shape variations, indicating pronounced interactions between mechanical phonons and intracavity photons (supplementary video). Figure \ref{fig:4}(d) illustrates the spectral response in the vicinity of the second harmonic of the electrical resonance, $2f_e = 7.01$ MHz. At this frequency, the localized combs centered around $2f_e$ interferes with the downstream comb structure formed around the electrical resonance itself. This interaction gives rise to a dual-comb configuration characterized by a refined frequency spacing, with the secondary free spectral range, FSR$_2=11.39$ kHz, associated with half of the primary spacing.

\begin{figure}[t!]
	\centering
	\begin{subfigure}[b]{0.94\textwidth}
		\centering
		\includegraphics[width=\textwidth]{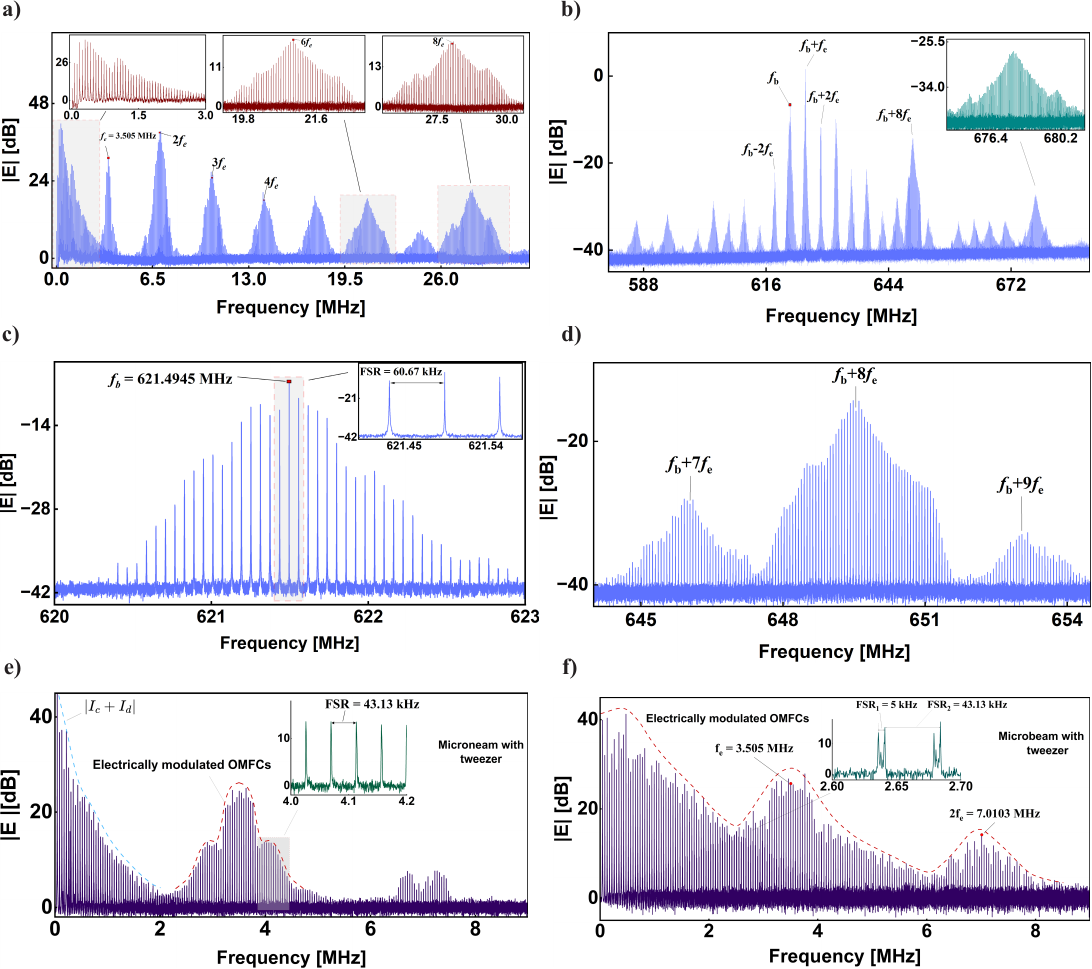}
		\vspace{0.2mm}
	\end{subfigure}
	\caption{Measured optical spectra. Response spectra of cavities based on curved microbeam variant \uppercase\expandafter{\romannumeral1}, pumped at the three-eighths span under 7 mTorr vacuum, yielding a localized cavity with resonance frequency $f_c=23.86$ THz. (a) Low-frequency spectrum showing half of the OMFCs centered at zero with FSR = 60.67 kHz, corresponding to the optically tuned first anti-symmetric bending mode, along with additional combs at $f_e=3.505$ MHz and harmonics. (b) High-frequency spectrum. (c) OMFCs centered at the carrier $f_b=621.495$ MHz. (d) Zoomed spectrum showing electrically modulated FCs at $f_b+7f_e$, $f_b+8f_e$, and $f_b+9f_e$. (e, f) Response spectra of cavities from microbeam variant \uppercase\expandafter{\romannumeral2}, exhibiting the formation of dual FCs with FSR$_1=5$ kHz and FSR$_2=43.13$ kHz, associated with the second in-plane bending mode\textquotesingle s oscillation frequency.}
	\label{fig:5}
\end{figure}

The high-frequency domain of the measured response is presented in Figure \ref{fig:4}(e). The spectrum extends over a broader frequency range spanning from 618 MHz to 627 MHz, as compared to the results shown in Figures \ref{fig:3}(f) and \ref{fig:5}(c). It exhibits optomechanical soliton FCs with a comb spacing of FSR=22.78 kHz. Following post-photodetection electrical modulation, a densely packed frequency comb emerges at the shifted center frequency $f_b+f_e$, preserving the same frequency spacing. The spectral components of this comb exhibit partial overlap with the soliton spectrum centered at $f_b$, resulting in the formation of a dual-comb structure extending across the megahertz frequency range. Notably, all observed combs vanish when the experiment is performed in air. This disappearance further confirms the formation of localized microcavities under reduced mechanical losses. The vacuum environment facilitates stimulated emission of coherent phonons to surpass overall mechanical losses at lower pump powers, a critical requirement for phonon lasing to emerge under blue-detuned optical pumping.

An optomechanical cavity with resonance frequency of approximately $f_c=23.86$ THz is achieved by positioning the laser spot at three-eighth length of the curved microbeam. Laser absorption within the microstructure\textquotesingle s material generates a thermo-mechanical force acting along the direction of the microbeam\textquotesingle s motion, thereby enhancing its transverse curvature. This simultaneously induces a tensile axial force as a consequence of the clamped–clamped boundary conditions. The combined influence of the increased curvature and the thermo-mechanically induced axial tension results in an upward shift of the natural frequency associated with the first anti-symmetric in-plane bending mode. The measured response spectrum displayed in Figures \ref{fig:5}(a)-\ref{fig:5}(d) demonstrates the amplification of phase-coherent and self-sustained mechanical oscillations, concurrently accompanied by an increase in intracavity field intensity, measured under the pump power same as previous.

A comparison between Figures \ref{fig:5}(a) and \ref{fig:3}(d) indicates that the kilohertz-range spectral lines span a broader frequency domain. This spectral broadening results in the overlap and eventual merging of the optomechanically induced comb lines with those originating from electrically modulated OMFCs centered at the electrical resonance frequency $f_e$. Further, the localized FCs centered at higher-order harmonics of the electrical resonance exhibit densely populated spectral structures, indicating the onset of combs\textquotesingle\ teeth interference. The high-frequency range of the optical wavepacket spectrum composed of Doppler-shifted electromagnetic waves and intracavity optical field is presented in Figure \ref{fig:5}(b). Similar to the results shown in Figure \ref{fig:3}(e), a broad comb structure is distributed around the central frequency $f_b$ with a spacing syntonized to the electrical resonance. The localized wavepackets formed in the vicinity of of each harmonic $f_b\pm nf_e\ (n=1,2,\ldots,9)$ exhibits improved spectral densities as compared to those shown in Figure \ref{fig:3}(e). The progressive decay in spectral line power, forming soliton-like envelopes around each localized FC, is indicative of cavity operation within the unresolved sideband regime. Figure \ref{fig:5}(c) shows a zoomed view of the spectrum in the vicinity of the carrier frequency ($f_b$). The spacing of the OMFCs is determined FSR=60.67 kHz corresponding to the natural frequency of the microbeam\textquotesingle s tuned second in-plane bending mode. Figure \ref{fig:5}(d) illustrates a comb structure centered at $f_b+8f_e$, characterized by prominent comb lines well above the noise floor, flanked by two weaker frequency combs located at $f_b+7f_e$ and $f_b+9f_e$. This spectral arrangement signifies the onset of overlap among the three comb families, a condition that can facilitate the formation of dual-frequency combs with enhanced spectral resolution.

The optical response spectra for an optomechanical cavity composed of variant \uppercase\expandafter{\romannumeral2} of the curved microbeam, with its second in-plane bending mode\textquotesingle s natural frequency at $f_2^{ib}\approx39$ kHz are presented in Figures \ref{fig:5}(e) and \ref{fig:5}(f). The cavity is pumped at two adjacent locations near the quarter-span of the microbeam, using the same laser power and frequency as in the previous experiments. Results indicate that the strength of the optomechanical interaction is highly sensitive to the spatial position of the excitation. Figure \ref{fig:5}(e) shows the corresponding response spectrum is primarily dominated by low-frequency components with an FSR of 43.13 kHz, extending up to 2 MHz, along with electrically modulated OMFCs centered at $f_e$. A slight shift in the laser spot position within the quarter-span vicinity produces a cavity with significantly enhanced coupling between the optical and mechanical domains, Figure \ref{fig:5}(f). The results reveal the generation of dual FCs with finer spectral spacing of 5 kHz.

\section{Conclusions}\label{sec13}
We investigated the emergence and dynamics of localized Fabry-P\'{e}rot cavity optomechanical resonators within MEMS architectures. The experimental results presented in Figures \ref{fig:3}-\ref{fig:5} highlight the significant potential risk associated with the unintended formation of optomechanical resonators within electrostatic MEMS structures. Specifically, when two reflective electrodes form a MEMS capacitor, optical interrogation can inadvertently create an optomechanical cavity. In this regime, the dominant optical response originates from intracavity fields strongly coupled to mechanical phonons, rather than from the Doppler-shifted light expected in conventional measurements. These findings emphasize the importance of careful laser spot positioning and appropriate optical filtering to avoid unintentional cavity formation and misinterpretation of experimental data.

We showed that optical pumping of the capacitive gap in an electrostatic MEMS device without electrical excitation establishes strong photon-phonon interactions, effectively generating distributed optomechanical resonators along the microbeam\textquotesingle s length. Our experimental results unveiled that the measured optical signal simultaneously encodes two critical contributions: (i) Doppler-shifted light, directly representing the mechanical motion of the microbeam, and (ii) intracavity optical field arising from the nonlinear optomechanical dynamics. Critically, when the probe laser partially illuminates the capacitive gap between actuation and movable electrodes, the combination of these contributions leads to significantly measurable deviations in the extracted MEMS motion, underscoring the necessity of accounting for optomechanical dynamics in optical measurements.

Further, we employed this system to establish and validate an efficient methodology for generating optomechanical soliton FCs using localized Fabry-P\'{e}rot cavities within MEMS architecture. By optically pumping the capacitive gap at various positions along the microbeam span, we demonstrated the generation of OMFCs with distinct FSR, each corresponding to the natural frequency of a certain mechanical mode. Remarkably, optical soliton wavepackets with a spectral line spacing of 5 kHz were realized using a single-wavelength CW laser at a modest pump power of 4 mW, without requiring integration of optical fibers into the MEMS.

Collectively, these results both highlight the critical influence of optomechanical interactions on conventional optical MEMS measurements and establish a practical, efficient platform for on-chip generation of optomechanical soliton frequency combs, advancing the integration of MEMS and optomechanical photonic technologies.

\section{Materials and methods}\label{sec11}
\subsection{Experimental validation}
A green, single-wavelength CW laser with a wavelength of 532 nm ($f_0=564$ THz) and optical power of 4 mW, emitted from a UHF Polytec LDV, is directed toward the microbeam to pump the microcavity, Figure \ref{fig:2}. The laser spot is aligned to partially occupy the in-plane gap between the movable and stationary mirrors, while also partially impinging on the sidewall of the movable micromirror. The fraction of incident light that enters the cavity undergoes multiple reflections between the two mirrors, generating counter-propagating intracavity photon pairs. These photon pairs exert radiation pressure on the suspended mirror along the in-plane (z-axis) direction, effectively acting as a driving force that induces mechanical displacement within the structure. The mechanical oscillations modulate the effective cavity length, leading to dynamic shifts in the cavity\textquotesingle s optical resonance frequency through parametric excitation, forming a motion-induced nonlinear back-action on the optical domain. This interaction gives rise to an in-plane cavity optomechanical resonator triggering phonon lasing under blue detuning.

Following this scenario, the light reflected toward the photodetector integrated into the LDV sensor head comprises two components: a Doppler-shifted electromagnetic wavepacket encoding information about the mechanical motion, $E_d(t)$, and a scattered intracavity optical field modulated at the mechanical oscillation frequencies, $E_c(t)$. The latter originates from optomechanical dynamics, wherein the motion-induced fluctuations in the cavity\textquotesingle s resonance frequency mediate the coupling between the mechanical phonons and intracavity photons. This configuration simultaneously facilitates optical actuation of the microcavity and enables optical measurement of the system response, thereby introducing a fully integrated optomechanical interrogation scheme, Figure \ref{fig:2}(a).

The optical forcing generated locally along the z-direction acts as an autonomous excitation, with its magnitude described by the instantaneous intracavity photon population. This force can effectively excite the in-plane bending modes of the flexible mirror (curved microbeam). The backscattered electromagnetic field directed to the sensor head of the LDV undergoes heterodyne demodulation, referenced to the source laser operating at frequency $f_0$, thereby retrieving the envelope of the corresponding optical wavepackets. The photodetector consequently produces an electrical signal at the modulated carrier frequency $f_b$ (Bragg cell-induced frequency shift), referred to as the photodetected field, expressed as follows.
\begin{align}
	E &= E_d+E_c \nonumber \\
	E_d &= \frac{1}{2}\left(\sum_{k=0}^{P}{A_ke^{ik\omega_mt}+cc}\right)e^{i2\pi f_b t}+cc \nonumber \\
	E_c &= \frac{1}{2}\left(\sum_{k=0}^{N}{B_ke^{ik\omega_mt}+cc}\right)e^{i2\pi f_b t}+cc, \quad N\gg P \label{eq:1a}
\end{align}
where, $\omega_m$ is the mechanical oscillation frequency. The acquired signal is transmitted to the LDV\textquotesingle s DMS for a post-photodetection demodulation processing. Within the DMS, a numerical in-phase and quadrature (I/Q) demodulation scheme, functioning as a secondary demodulation stage, is implemented to retrieve the photodetected field envelope. This procedure simultaneously enables calibration of the signal into displacement and velocity quantities. Due to the slower dynamics of this numerical demodulation process relative to the initial photodetection stage, the demodulated output may retain residual components of the photodetected signal, $E_{res}$. As a result, the post-processed output signal encapsulates spectral content spanning both low-frequency components, extracted via I/Q demodulation, and high-frequency components in the hundreds of megahertz range, originating from initial photodetection. It is important to note that the I/Q demodulated spectrum itself encompass contributions from two distinct sources: 1) Doppler-shifted electromagnetic waves encoding information about mechanical motions of the suspended mirror, and 2) modulated intracavity optical field originating from the optomechanical dynamics developed within the localized in-plane microcavity.
\begin{align}
	E_{I/Q} &= \sum_{k=0}^{N}{(A_k+B_k)e^{ik\omega_mt}+cc} + E_{res}\ , \ (A_k=0,\; k>P) \label{eq:1b}
\end{align}

When the laser pumping is inactive, inspecting the spectrum of the post-processed signal reveals the presence of a modulated electrical resonance at $f_e=3.505$ MHz, attributable to the vibrometer\textquotesingle s oscilloscope circuitry, Figure \ref{fig:3}(a). Additionally, a persistent spectral component at $f_b=621.495$ MHz is observed, corresponding to residual electrical signals generated by the photodetector. This high-frequency component vanishes when the transmission line carrying the photodetected signal is disconnected from the oscilloscope. Activating the laser and directing it toward a reflective stationary surface amplifies the component at the carrier frequency $f_b$, which subsequently undergoes modulation by $f_e$, \ref{fig:3}(b). This behavior is attributed to an electrical quadratic ($\chi^2$) nonlinearity inherent in the oscilloscope circuitry, leading to the generation of equidistant FCs centered around $f_b$. Also, integer harmonics of $f_e$, up to the eleventh order, appear within the low-megahertz range of the spectrum, reflecting the nonlinearity\textquotesingle s back-action mechanism. Notably, these measurements were performed in the absence of the microcavity. Throughout this section, the electrically modulated post-processed signal is referred to as the system response, denoted $E$.

\subsection{Mechanical characterization}
The cavity\textquotesingle s mechanical modes were characterized using electrostatic excitation in air. Activating potential difference between the movable (curved microbeam) and the stationary mirrors within pulse excitation extracts the first few in-plane bending modes of the curved microstructure. Using a pulse excitation with amplitude of 15 V, pulse frequency of 25 Hz, and duty cycle of $0.1\%$, the following mechanical modes were determined: the first in-plane bending at $f_1^{ib}=33.13$ kHz, second in-plane bending at $f_2^{ib}=56.78$ kHz, and the third in-plane bending at $f_3^{ib}=111.8$ kHz, modes along with the first out-of-plane bending mode at $f_1^{ob}=67.38$ kHz, Figure 3(b). The natural frequency of the second (first anti-symmetric) in-plane bending mode is nearly twice that of the first symmetric in-plane mode, $f_2^{ib}\approx2f_1^{ib}$. Further, the natural frequency of the microbeam\textquotesingle s third (second symmetric) in-plane bending mode is nearly twice that of the second in-plane bending mode, $f_3^{ib}\approx2f_2^{ib}$. 

\vspace{3mm}

\backmatter

\bmhead{Supplementary Information}

This article has accompanying supplementary materials and video.

\bmhead{Author Contribution}

SR: Investigation, concept, methodology, formal analysis, data curation, software, writing original draft, experimental investigation, numerical simulations, and reviewing and editing;

\bmhead{Funding}

The authors did not receive support from any organization for the submitted work.









\bibliography{sn-bibliography}

\end{document}


\title[In-plane cavity optomechanics]{\centering Supplementary Materials\\ Localized In-Plane Cavity Optomechanics in MEMS}


\author[]{\fnm{Sasan} \sur{Rahmanian}}\email{s223rahm@uwaterloo.ca}

\affil[]{\orgdiv{Systems Design Engineering Department}, \orgname{University of Waterloo}, \orgaddress{\street{200 University Ave West}, \city{Waterloo}, \postcode{N2L 3G1}, \state{Ontario}, \country{Canada}}}

\maketitle

\section{Mathematical model}\label{sec1}

The optomechanical cavity illustrated in Figure 1(b) consists of two mirrors: a fixed side mirror, and a curved microbeam acting as the suspended mirror, fabricated in-plane on a SOI wafer. For a relatively large in-plane displacement $w(x;t)$, the optical resonance frequency of a localized cavity formed at position $x_0$ from the origin $O$ can be written as $\omega_{opt}=\frac{\pi c}{g_0-h/2+w_0(x_0)+w(x_0;t)}$. The total optical energy $U_{opt}$ stored in the cavity can be written in terms of the complex-valued and slowly varying envelope $\alpha$ of the optical mode, which is normalized such that $|\alpha|^2$ represents the number of instantaneous intracavity photons. The expression for the optical energy stored in a cavity formed at position $x_0$ is $U_{opt}=\hbar\ \omega_{opt}|\alpha|^2$, where $\hbar$ is the reduced Planck constant. The radiation pressure force applied to the structure at $x=x_0$ is: $F_{opt}(w;t)=-\partial U_{opt}/\partial w$ introducing a lumped displacement-dependent optical excitation for the curved microbeam. The suspended microstructure can move in response to the optical pumping, effectively creating an optical microcavity, wherein light can bounce back and forth between the two mirrors. The equations of motion describing the cavity\textquotesingle s autonomous dynamics encompassing optomechanical coupling between the optical and the mechanical domains can be expressed as follows \cite{miri2018optomechanical, wan2025optomechanical}.
\begin{equation}\label{eq1}
	\begin{aligned}
		\frac{d\alpha}{dt} &= 
		\left( i\big( \omega_L - \omega_{opt}\big) - \tfrac{\kappa}{2} \right)\alpha 
		+ \sqrt{\kappa_e}\, s_{in}, \\[6pt]
		\rho A \ddot{w} + \tilde{c}_d \dot{w} + EI w'''' 
		&- \left[\widetilde{N} + \tfrac{EA}{2L} \int_{0}^{L} \left( {w'}^2 + 2 w' w_0' \right)\, dx \right]
		\left( w'' + w_0'' \right) \\
		&= \frac{\pi c \hbar |\alpha|^2}{\left(g_0 - \tfrac{h}{2} + w_0 + w\right)^2}\, \delta(x-x_0), \\[6pt]
		\text{BCs: } & \quad w = w' \quad \text{at} \quad x=0,\, L
	\end{aligned}
\end{equation}
where, $\omega_L$ is the laser pump frequency, and $\kappa=\kappa_e+\kappa_l$ represents the sum of external and internal optical losses, $s_{in}$ is the input photon flux. $\widetilde{N}$ denotes the axial load applied to the microbeam by the supports used for modelling the influence of residual axial stress. $A$ and $I$ express the cross-sectional area and second moment of area of the microbeam. $\tilde{c}_d$ is the viscous damping coefficient, and $\delta$ is the Dirac Delta function with unit of m$^{-1}$ centered at $x_0$. Defining the following dimensionless parameters,
\begin{equation}\label{eq2}
	\begin{aligned}
		\widetilde{w}= \frac{w}{g_0},\ \ 
		\widetilde{w}_0 = \frac{w_0}{g_0},\ \
		\widetilde{\delta} = g_0 \delta,\ \ 
		\widetilde{x}= \frac{x}{L}, \ \
		\widetilde{t}= \frac{t}{t^\ast},\ \
		\text{where} \quad t^\ast = \sqrt{\frac{\rho A L^4}{EI}}.
	\end{aligned}
\end{equation}
and substituting them into Eq. (\ref{eq1}), and eliminating the tildes for the sake of simplicity, the non-dimensional optomechanical equations of motion are obtained as:
\begin{equation}\label{eq3}
	\begin{aligned}
		\frac{d\alpha}{dt} &= \left(i\left(\omega_l - \frac{\beta_1}{1 - R + w_0 + w}\right) - \chi_1\right)\alpha + \chi_2, 
		&\quad & x = x_0, \\[1mm]
		\ddot{w} + c_d \dot{w} + w^{\prime\prime\prime\prime} 
		&- \left[N + \beta_2 \int_{0}^{1} \left({w^\prime}^2 + 2 w^\prime w_0^\prime \right) dx \right] \left(w^{\prime\prime} + w_0^{\prime\prime}\right)\\&= \beta_3 \frac{|\alpha|^2}{(1 - R + w_0 + w)^2} \, \delta(x - x_0), \\
		&\text{BCs:} \quad w = w^\prime = 0 \quad \text{at} \quad x = 0, 1.
	\end{aligned}
\end{equation}
where,
\begin{equation}\label{eq4}
	\begin{aligned}
		& \omega_l = t^\ast \omega_L, \quad 
		\beta_1 = \frac{t^\ast \pi c}{g_0}, \quad 
		\chi_1 = \frac{\kappa t^\ast}{2}, \quad 
		\chi_2 = \sqrt{\kappa_e}\, s_{in} t^\ast, \\[1mm]
		& c_d = \frac{t^\ast \tilde{c}_d}{\rho A}, \quad 
		N = \frac{L^2 \widetilde{N}}{EI}, \quad 
		R = \frac{h}{2 g_0}, \quad 
		\beta_2 = \frac{A g_0^2}{2 I}, \quad 
		\beta_3 = \frac{\pi c \hbar L^4}{EI\, g_0^4}.
	\end{aligned}
\end{equation}
In Eq. (\ref{eq4}), $\omega_l$ and $\omega_c$, respectively, represent the non-dimensional laser pump frequency and the cavity\textquotesingle s optical resonance frequency. The displacement function $w$ is expanded in terms of the eigenfunctions of the corresponding linear undamped clamped-clamped microbeam ($\varphi_i$), $w=\sum_{i=1}^{M}{q_i(t)\varphi_i(x)}$, where $q_i$s represent the mechanical mode\textquotesingle s generalized coordinates. Substituting this expansion into Eq. (\ref{eq3}) for $M=7$, applying the Galerkin method within the interval of [0,\ 1], leads to $M+1$ nonlinearly coupled ordinary differential equations (ODEs) $q_i$s and the complex-valued optical mode, $\alpha$.
\begin{equation}\label{eq5}
	\begin{aligned}
		 \frac{d\alpha}{dt} &= i\Bigg(\omega_l - \frac{\beta_1}{1 - R + w_0 + \sum_{i=1}^{M} q_i \varphi_i} + G_{th} T \Bigg) \alpha - \chi_1 \alpha + \chi_2, \quad x = x_0, \\[1mm]
		 \ddot{q}_n + & c_d \dot{q}_n + \omega_n^2 q_n 
		- \int_0^1 \varphi_n \Bigg[ \Bigg( N + \beta_2 \int_0^1 \Big( \big(\sum_{i=1}^{M} q_i \varphi_i^\prime \big)^2 + 2 w_0^\prime \sum_{i=1}^{M} q_i \varphi_i^\prime \Big) dx \Bigg)\\
		&\Big( \sum_{i=1}^{M} q_i \varphi_i^{\prime\prime} + w_0^{\prime\prime} \Big) \Bigg] dx = \beta_3 \frac{|\alpha|^2 \varphi_n(x_0)}{\left(1 - R + w_0(x_0) + \sum_{i=1}^{M} q_i \varphi_i(x_0)\right)^2} + \mu T, \\[1mm]
		& \dot{T} = -\gamma_{th} T + \beta_4 |\alpha|^2.
	\end{aligned}
\end{equation}

Equation (\ref{eq4}) incorporates the opto-thermal coupling arising from intracavity photon absorption by both the mechanical resonator and the cavity medium, whereby a fraction of the optical energy is converted into heat. Absorption within the mechanical resonator leads to a temperature rise identified by its heat capacity and thermal relaxation, producing a thermo-mechanical force $\mu T$ directed along the resonator\textquotesingle s oscillation. Further, the thermo-optic pathway stems from heating-induced modifications of the cavity\textquotesingle s optical path length, as the refractive index of the cavity medium varies with temperature, thereby shifting the optical resonance frequency \cite{wan2025optomechanical}. Here, $T(t)$ denotes the temperature rise relative to equilibrium. The third dimensionless equation in Eq. (\ref{eq5}) describes the temperature dynamics of the mechanical structure induced by opto-thermal coupling. The parameters $G_{th}$, $\mu$, and $\gamma_{th}$, represent, respectively, the thermo-optic coupling constant, the thermo-mechanical force coefficient, and the thermal decay rate, while $\beta_4$ encapsulates the combined effects of heat capacity and total optical absorption.

The mechanical oscillator\textquotesingle s dynamics incorporate the mid-plane stretching resulting from large bending deformation under clamped-clamped boundary conditions, along with the microbeam\textquotesingle s initial curvature. These structural nonlinearities, together with the optomechanical coupling induced by the fractional and displacement-dependent optical radiation pressure, source the potential nonlinearities in the mechanical domain. Further, the optomechanical coupling, characterized by the dependence of the cavity resonance frequency on mechanical motions, expressed as $\omega_c(w)\alpha$, introduces modal interaction in the optical domain. Unlike optical microring resonators, where soliton formation relies critically on Kerr nonlinearity and optical dispersion, optomechanical resonators can generate optical soliton response without the prerequisite of these optical effects. The terms $\ddot{w}$ and  $w''''$ are representative for mechanical dispersion, which are potentially capable of counterbalancing the system's nonlinearities, leading to the emergence of optical soliton wavepackets.

\section{Static solution and cavity\textquotesingle s eigenvalue problem}\label{sec2}

The static optomechanical responses are determined by dropping the time-derivative terms in Eq. (\ref{eq3}). Those responses represent the static equilibrium of the nonlinear autonomous dynamics expressed in Eq. (\ref{eq3}), denoted ($\alpha^\ast,w^\ast$). The displacement function $w^\ast$ is expanded in terms of the eigenfunctions of the linear undamped clamped-clamped microbeam corresponding to the mechanical resonator, $\varphi_i$, $w^\ast=\sum_{i=1}^{M}{c_i\varphi_i}$, where $c_i$s denote the modal constants. Substituting this expansion into Eq. (\ref{eq3}) for $M=7$, multiplying both sides of the equation by $\varphi_n$, and taking integral from the resulted equations within the interval of [0,\ 1], leads to $M+1$ nonlinearly coupled algebraic equations on $c_i$s and the complex-valued optical mode, $\alpha^\ast$. This set of equations are solved using nonlinear solvers in MATLAB.

The optomechanical eigenvalue problem is formulated via linearizing Eq. (\ref{eq3}) around the obtained equilibrium point. The system dynamic responses are perturbed around ($\alpha^\ast,w^\ast$), as follows:
\begin{equation}\label{eq6}
	\begin{aligned}
		& \alpha(t) = \alpha^\ast(x) + \alpha_d(t), \\
		& w(x;t) = w^\ast(x) + w_d(x;t).
	\end{aligned}
\end{equation}
Substituting Eq. (\ref{eq6}) into Eq. (\ref{eq3}), expanding the cavity\textquotesingle s resonance frequency, $\omega_c$, and the expression for the radiation pressure using Taylor series expansion around $(\alpha_d,w_d)=(0,0)$ keeping only the linear terms, and eliminating the time-independent terms in the optical and mechanical domains by satisfying the static equations, yields,
\begin{subequations}\label{eq7}
	\begin{align}
		\dot{\alpha}_d &= 
		\left(i\left(\omega_l - \frac{\beta_1}{1-R+w_0+w^\ast}\right) - \chi_1 \right) \alpha_d
		+ \frac{i \alpha^\ast \beta_1}{\left(1-R+w_0+w^\ast\right)^2} w_d,\ \label{eq7a} \\ \label{eq7b}
		\ddot{w}_d &+ c_d \dot{w}_d + w_d^{\prime\prime\prime\prime} 
		- w_d^{\prime\prime} \left[N + \beta_2 \int_0^1 \left( 2 {w^\ast}^\prime w_0^\prime + {{w^\ast}^\prime}^2 \right) ds \right] \notag \\
		&\quad - \beta_2 \left( {w^\ast}^{\prime\prime} + w_0^{\prime\prime} \right) 
		\int_0^1 \left( 2 w_0^\prime w_d^\prime + 2 {w^\ast}^\prime w_d^\prime \right) ds \notag \\
		&= \beta_3 \left[ \frac{\alpha^\ast \bar{\alpha}_d + \bar{\alpha}^{\ast} \alpha_d}{\left(1-R+w_0+w^\ast\right)^2} 
		- \frac{2 |\alpha^\ast|^2 w_d}{\left(1-R+w_0+w^\ast\right)^3} \right] \delta(x-x_0) 
	\end{align}
\end{subequations}

\begin{figure}[h]
	\centering
	\begin{subfigure}[b]{0.94\textwidth}
		\centering
		\includegraphics[width=\textwidth]{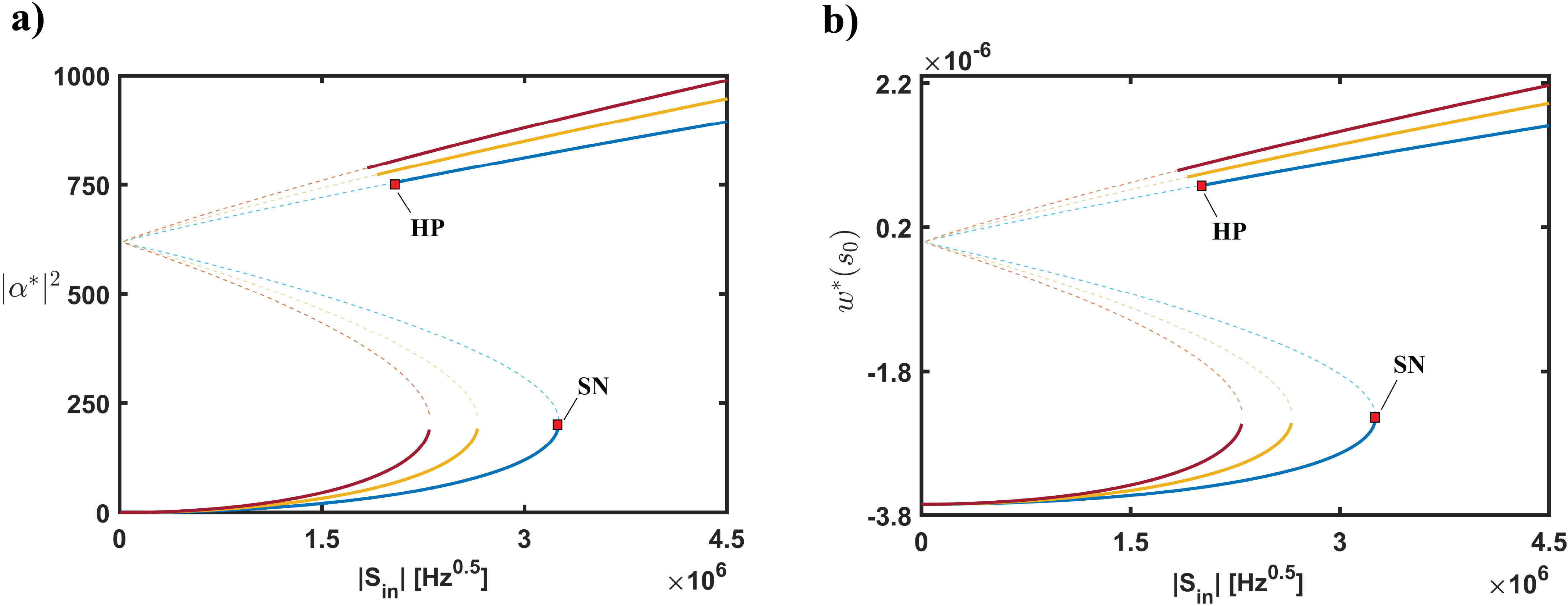}
		\vspace{0.1mm}
	\end{subfigure}
	\caption{Stationary cavity responses. (a) Intracavity instantaneous photon number and (b) displacement of the curved movable mirror as functions of the input photon flux for a localized cavity at position $s_0=0.25$. The optical loss coefficients are set: $\kappa=2$ MHz (blue), $\kappa=3$ MHz (yellow), and $\kappa=4$ MHz (red). Solid lines represent stable solution branches, while dotted lines indicate unstable solutions. The simulation was performed assuming a non-dimensional axial force $N=-0.025$.}
	\label{fig1}
\end{figure}

To establish the system\textquotesingle s eigenvalue problem using Eq. (\ref{eq7}), the partial differential equation (PDE) is discretized to a set of seven second-order ODEs. This is achieved by expanding the dynamic displacement component $w_d$ as a linear combination of the eigenfunctions $\varphi_i$, $w_d=\sum_{i=1}^{M=7}{q_i\varphi_i}$, and applying the Galerkin method. Here, $q_i$ denotes the generalized coordinate associated with the $i^{th}$ mechanical mode. This system of equations are then converted into a $2n$-space by defining the state vector $|x\rangle=\{q_1,q_2,\ldots,q_7,{\dot{q}}_1,{\dot{q}}_2,\ldots,{\dot{q}}_7,\alpha_d\}$. This introduces fourteen first-order linear ODEs coupled to the single optical mode expressed in Eq. \ref{eq7a}. The fifteen first-order linearly coupled optomechanical ODEs can be written in the following matrix form.
\begin{equation}\label{eq8}
	\left| \dot{\mathbf{x}} \right\rangle = \mathbf{A} \left| \mathbf{x} \right\rangle
\end{equation}
By expanding the system response in modal form, $|x\rangle=e^{\lambda t}|X\rangle$, and substituting it into Eq. (\ref{eq8}), we obtain the optomechanical eigenvalue problem as:
\begin{equation}\label{eq9}
	\left( \mathbf{A} - \lambda \mathbf{I} \right) \left| \mathbf{X} \right\rangle = \mathbf{0}
\end{equation}
Equation (\ref{eq9}) results in global modes, each contributed by both the optical and mechanical degrees of freedom. For the parameter values listed in Table 1, the stationary responses of a localized cavity formed at $s_0=0.25$ are shown in Figure \ref{fig1} as functions of the input photon flux, evaluated for three values of the optical loss coefficient $\kappa$. At low input flux, both the intracavity photon number and the normalized microbeam displacement remain close to zero. As the optical excitation amplitude increases, these quantities grow until the equilibrium undergoes a saddle–node (SN) bifurcation at $|s_{in}|=3.2551\times10^6$ Hz$^{0.5}$. Beyond this threshold, the system exhibits abrupt jumps in its optomechanical steady-state responses to large-amplitude stable solution branches, which continue to increase with further pump enhancement. When the parameter is swept backward along this branch, the cavity undergoes a primary Hopf (HP) bifurcation at $|s_{in}|=2.0431\times10^6$ Hz$^{0.5}$.

\begin{figure}[t!]
	\centering
	\begin{subfigure}[b]{0.92\textwidth}
		\centering
		\includegraphics[width=\textwidth]{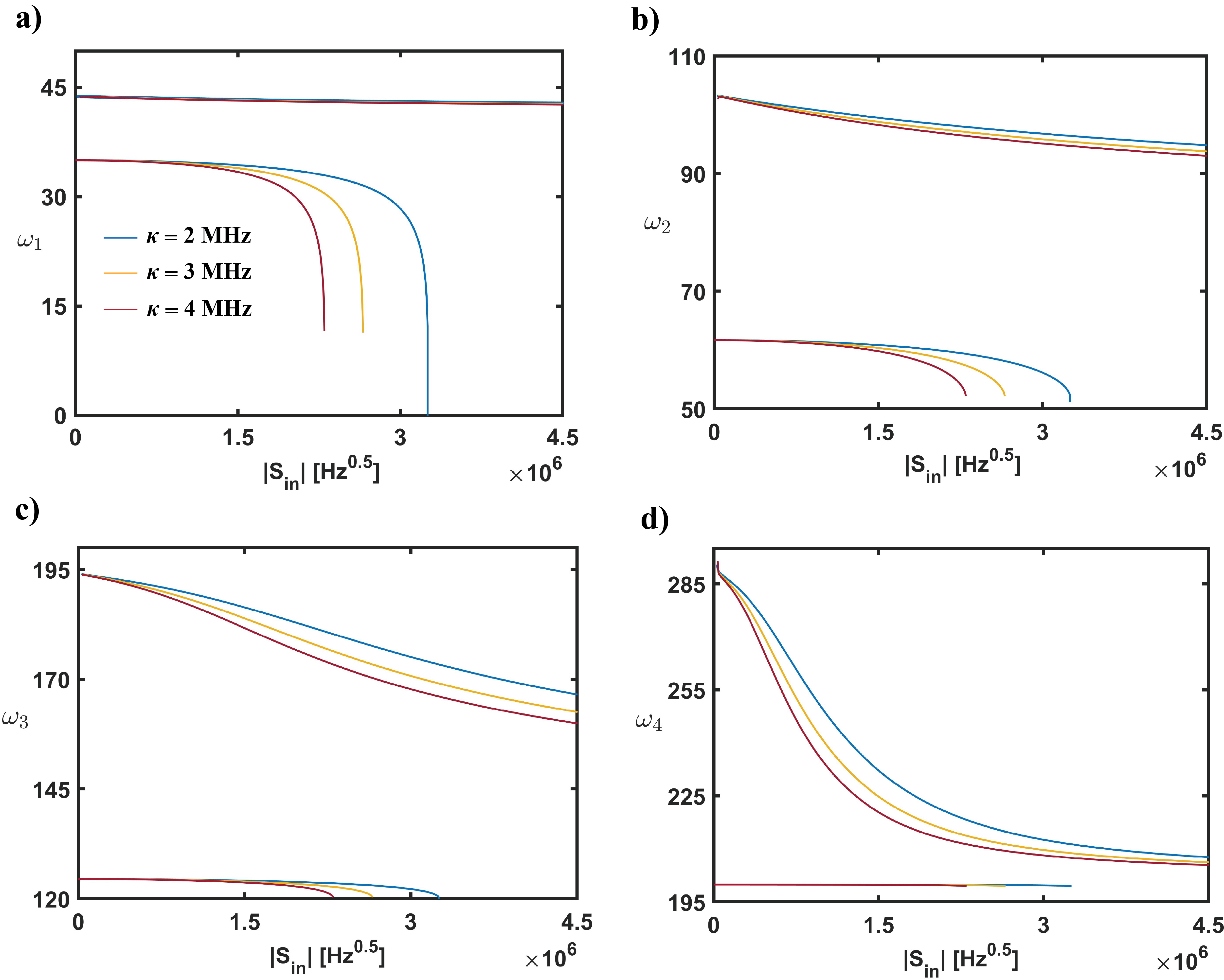}
		\vspace{0.1mm}
	\end{subfigure}
	\caption{Optomechanical natural frequencies of oscillations around the equilibria presented in Figure \ref{fig1}. The cavity\textquotesingle s (a) first, (b) second, (c) third, and (d) fourth natural frequency, for three values of the optical loss coefficient: $\kappa=2$ MHz (blue curve), $\kappa=3$ MHz (yellow curve), and $\kappa=4$ MHz (red curve).}
	\label{fig2}
\end{figure}

\begin{figure}[t!]
	\centering
	\begin{subfigure}[b]{0.89\textwidth}
		\centering
		\includegraphics[width=\textwidth]{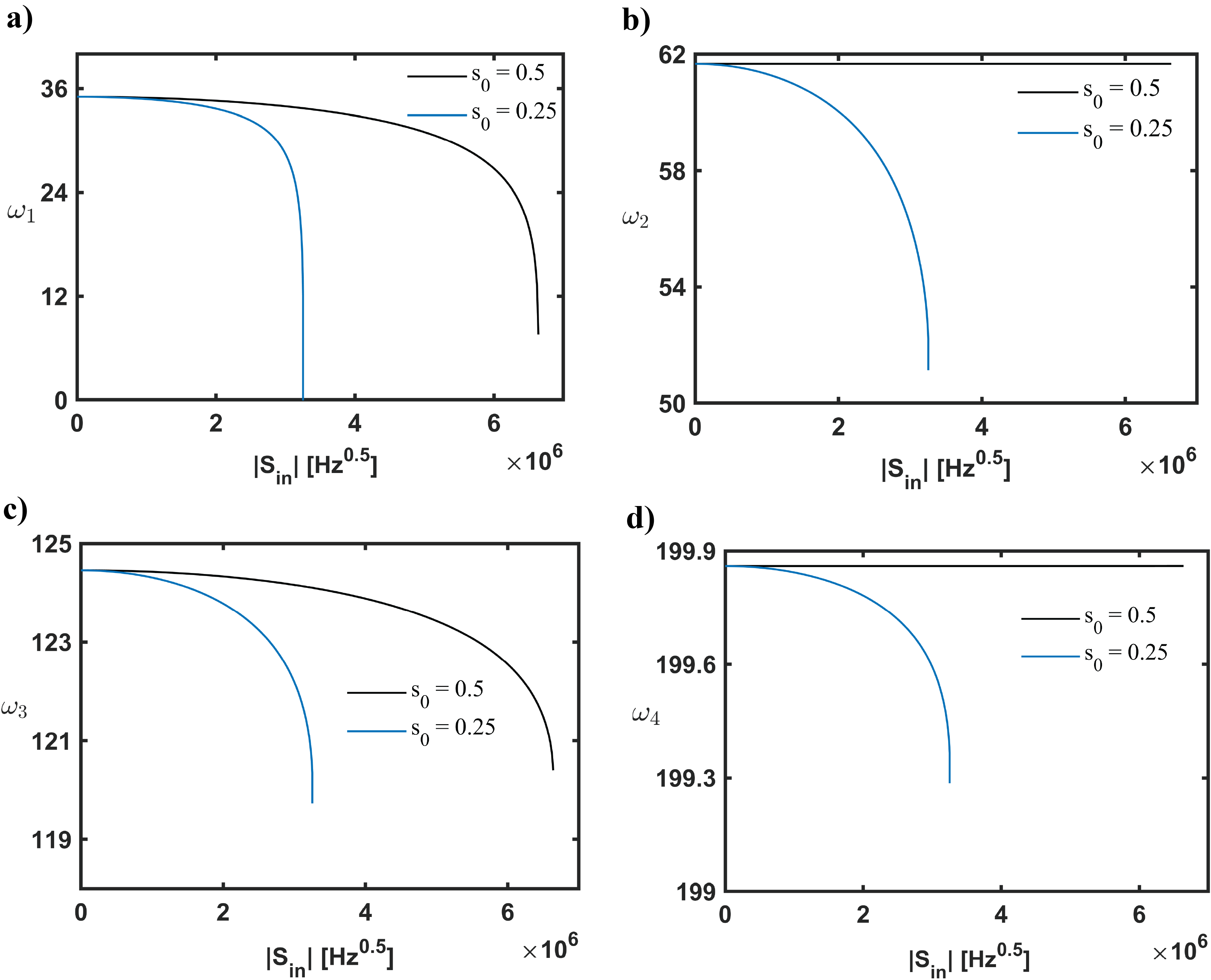}
		\vspace{0.1mm}
	\end{subfigure}
	\caption{The low-frequency branches of natural frequencies associated with two cavities localized at $s_0=0.25$ and $s_0=0.5$. The cavity\textquotesingle s (a) first, (b) second, (c) third, and (d) fourth natural frequency. The optical loss coefficient is $\kappa=2$ MHz.}
	\label{fig3}
\end{figure}

Figure \ref{fig2} exhibits the natural frequencies of the first four optomechanical modes versus input photon flux, corresponding to the static equilibria reported in Figure \ref{fig1}. The low-frequency branches represent oscillations about the low-amplitude stable equilibria, whereas the high-frequency branches correspond to the large-amplitude stationary states. Notably, the segments of the high-frequency branches located to the left of the HP bifurcation point indicate the natural frequencies at which perturbations around the unstable equilibria evolve along outward spiraling trajectories. The natural frequencies along the low-frequency branches decrease progressively with increasing input photon flux. In cavities with enhanced optical loss, these low-frequency branches emerge over restricted ranges of input flux. The low-frequency branches of the first four natural frequencies associated with two cavities localized at $s_0=0.25$ and $s_0=0.5$ are shown in Figure \ref{fig3}. Positioning the cavity  at the microbeam\textquotesingle s midspan, postpones the occurrence of SN bifurcation to $|s_{in}|=6.64001\times 10^6$ Hz$^{0.5}$, thereby extending the low-frequency branches over a broader range of the optical excitation amplitude. At the midspan location, which serves as a nodal point for the second and fourth mechanical bending modes, their spatial overlap with the intracavity optical field vanishes. Further, small static displacements induce negligible variations in the microbeam\textquotesingle s curvature. As a result, the natural frequencies associated with the second and fourth modes remain unchanged across the excitation range. However, the first and third mechanical modes, due to their strong coupling with the cavity mode, exhibit considerable variations across the excitation range.

\begin{table}[ht]
	\centering
	\caption{The optomechanical cavity properties.}
	\begin{spacing}{1.1}
		\begin{tabular}{|l|c|c|}
			\hline
			\textbf{Parameter}	& \textbf{Value}  \\
			\clineB{1-2}{2}
			Beam length, $L$ & 1000 \textmu m \   \\
			Beam width, $b$ & 9 \textmu m  \  \\
			Beam thickness, $h$ & 3 \textmu m \ \\
			Initial gap, $g_0$ & 10 \textmu m \  \\
			Midspan rise, $h_0$ & 3 \textmu m \  \\
			Mechanical damping coefficient, $c_d$ & 0.05 \  \\
			\hline
		\end{tabular}\label{tab1}
	\end{spacing}
\end{table}

\section{Dynamic response}\label{sec3}

\begin{figure}[t!]
	\centering
	\begin{subfigure}[b]{0.91\textwidth}
		\centering
		\includegraphics[width=\textwidth]{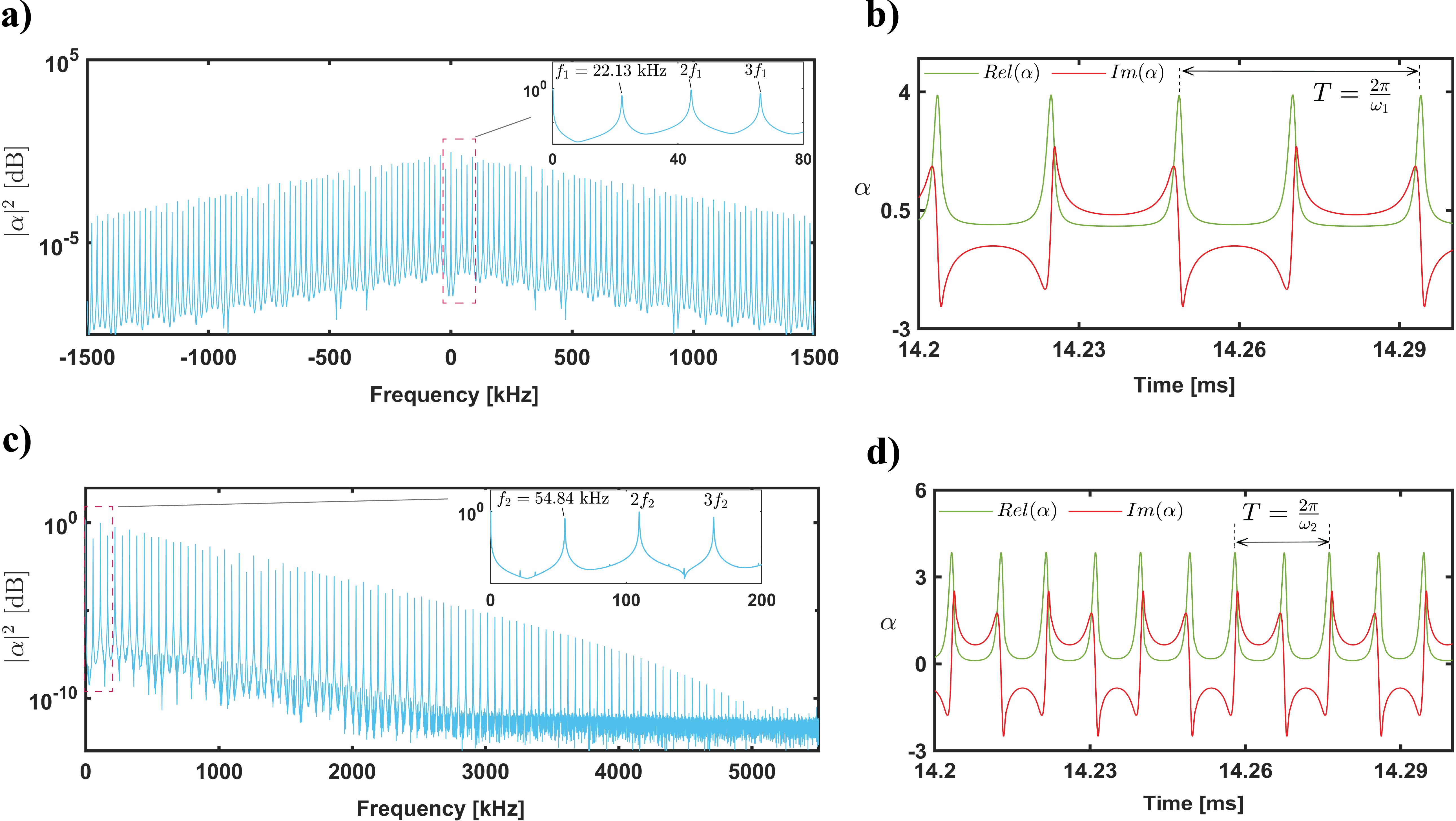}
		\vspace{0.1mm}
	\end{subfigure}
	\caption{Intracavity optical mode response obtained from numerical simulations for a curved microbeam with a midspan rise of $h_0=1$ \textmu m, an optical loss coefficient of $\kappa=10$ MHz, and an input photon flux of $|s_{in}|^2=81$ MHz. The spectral response is shown for: (a) a cavity localized at $s_0=0.5$, with the pump frequency blue-detuned by twenty times the fundamental mechanical mode, $\omega_l=\omega_c+20\omega_1$; and (c) a cavity positioned at $s_0=0.25$, with the pump laser frequency blue-detuned by ten times the second mechanical mode, $\omega_l=\omega_c+10\omega_2$. (b) and (d) depict the corresponding time-domain evolution of the real and imaginary components of the intracavity field amplitude.}
	\label{fig4}
\end{figure}

Figure \ref{fig4} illustrates the frequency contents and time-domain responses of the intracavity optical field, for a curved microbeam with a midspan rise of $h_0=1$ \textmu m. All other parameters are the same as those listed in Table 1. The optical loss coefficient and input photon flux are set $\kappa=10$ MHz and $|s_{in}|^2=81$ MHz. Positioning the local cavity at middle point of the mechanical resonator, $s_0=0.5$, and setting the optical pump frequency at $\omega_l=\omega_c+20\omega_1$, breaks the stability of the system equilibrium and leads to the formation of a periodic pulse train with a pulse time-spacing corresponding to the period of the microbeam\textquotesingle s first symmetric bending mode, $T_p=1/f_1$, Figure \ref{fig4}(b).
The optical response spectrum exhibits optomechanical FCs with FSR=$f_1$, Figure \ref{fig4}(a). In our experiments, the laser frequency remains stationary. To replicate the system dynamics under the conditions comparable to those in our experimental study, the pump frequency detuning is adjusted twenty times the natural frequency of the curved beam\textquotesingle s first in-plane bending mode, for an optical cavity localized at the structure\textquotesingle s midpoint. It is important to note that the midpoint of the microbeam coincides with a node of the anti-symmetric bending modes, rendering their contribution to the cavity resonance shift negligible. Among the symmetric bending modes, the first mode exhibits a maximal displacement at the midspan, resulting in the strongest spatial overlap with the intracavity optical field, and is consequently the predominant symmetric mode considered in the optical resonance shift. Figure \ref{fig4}(c) displays the optical response spectrum when the local cavity is positioned at one-quarter of the movable mirror\textquotesingle s length, $s_0=0.25$. In this configuration, the optical resonance frequency of the cavity increases relative to the previous case. Accordingly, the pump frequency is chosen as $\omega_l=\omega_c+10\omega_2$ to ensure that the system dynamics are simulated under pumping frequency comparable to those of the earlier scenario. At this cavity position, the microbeam\textquotesingle s first anti-symmetric bending mode exhibits its maximum amplitude, thereby establishing a dominant spatial overlap with the cavity\textquotesingle s electromagnetic field, while the fundamental harmonic appeared in the spectrum aligns with the natural frequency of this mode. The real and imaginary parts of the intracavity optical field are shown in Figure \ref{fig4}(d).

\begin{figure}[h]
	\centering
	\begin{subfigure}[b]{0.93\textwidth}
		\centering
		\includegraphics[width=\textwidth]{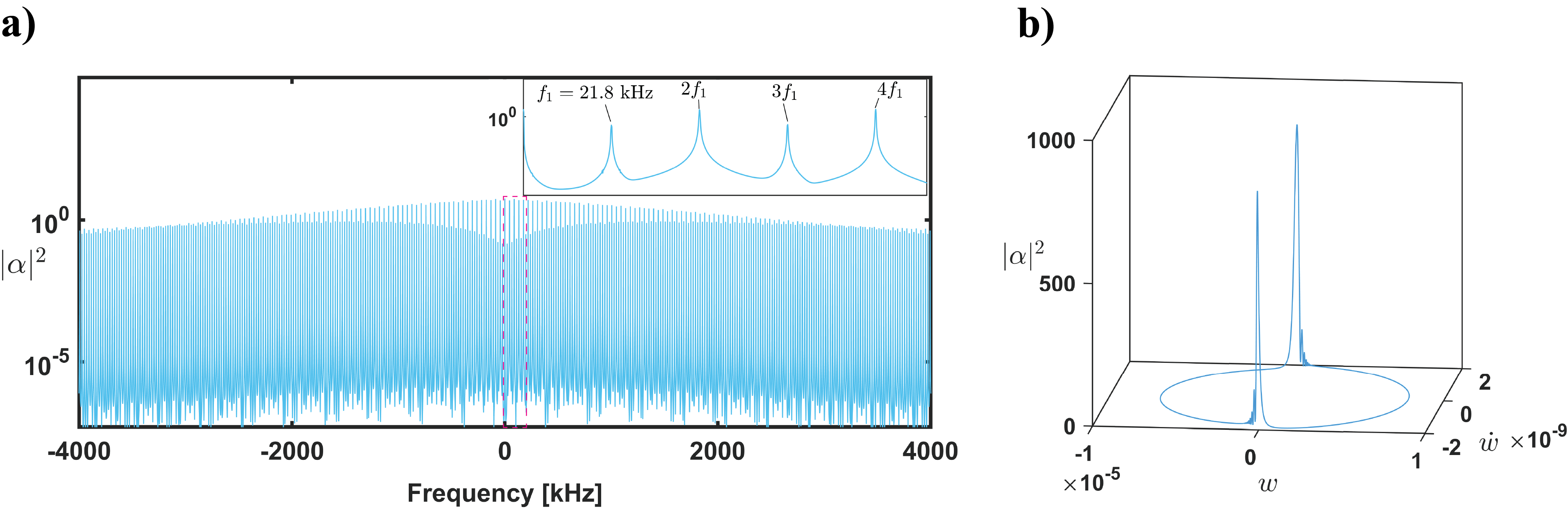}
		\vspace{0.1mm}
	\end{subfigure}
	\caption{Optical response of a cavity localized at the microbeam midspan ($s_0=0.5$) with an input photon flux of $|s_{in}|^2=8.1$ GHz. All other parameters are identical to those presented in the caption of Figure 7. (a) Frequency-domain response showing a dense optical wavepacket exhibiting a comb structure with free spectral range equal to the first mechanical resonance, FSR=$f_1=21.8$ kHz. (b) Corresponding system trajectory in the three-dimensional phase space, showing a periodic orbit with periodicity $T=1/f_1$.}
	\label{fig5}
\end{figure}

By enhancing the input photon flux to $|s_{in}|^2=8.1$ GHz while keeping all other parameters unchanged and localizing the cavity at the midpoint of the curved microbeam, $s_0=0.5$, a populated set of optomechanical FCs is produced. This sustains the elevated comb power across an expanded frequency range extending to 4 MHz, with a FSR=$f_1=21.8$ kHz, shown in Figure \ref{fig5}(a). Figure \ref{fig5}(b) presents a three-dimensional phase space of the cavity resonator, defined by the mirror\textquotesingle s displacement and velocity at $s_0$, together with the instantaneous intracavity photon number. The trajectory corresponds to a pulse-type periodic orbit, where the transition time between successive peaks equals half the period of the first bending mode.

\begin{figure}[t!]
	\centering
	\begin{subfigure}[b]{0.92\textwidth}
		\centering
		\includegraphics[width=\textwidth]{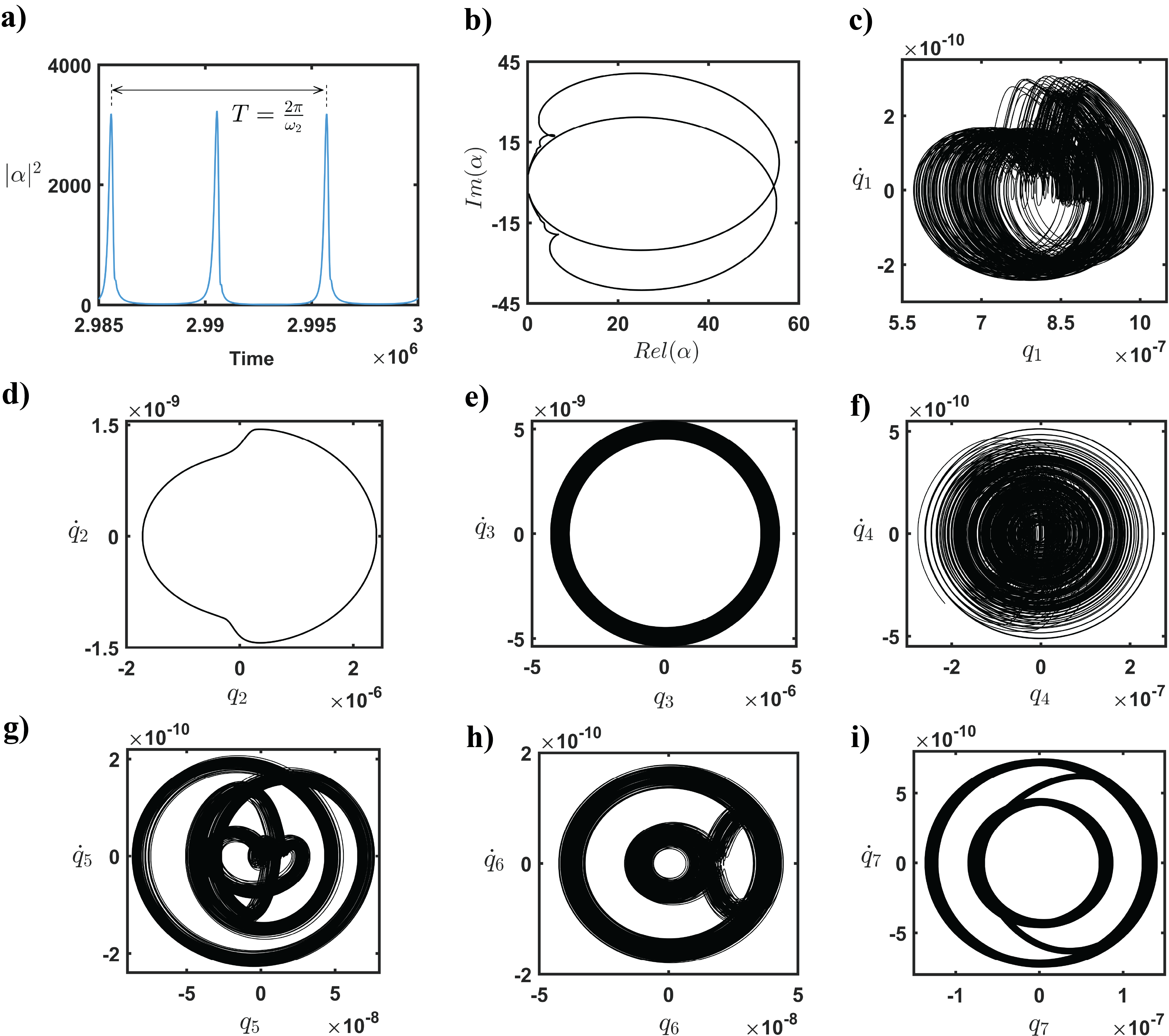}
		\vspace{0.1mm}
	\end{subfigure}
	\caption{The system optomechanical responses for a cavity localized at $s_0=0.375$, with an optical loss coefficient of $\kappa=22$ MHz and a pump intensity of $|s_{in}|^2=40$ GHz. (a) Time-domain evolution of instantaneous number of intracavity photons, showing a cnoidal wavepacket with a repetition rate equal to the second mechanical mode resonance. (b) Corresponding optical periodic orbit incorporated by the real and imaginary components of the optical field amplitude. (c)-(i) depict the displacement-velocity phase spaces associated with the first seven mechanical modes.}
	\label{fig6}
\end{figure}

For a cavity formed at $s_0=0.375$, the experimental results (Figure 5(a)-5(d)) indicate that the intracavity field is primarily modulated by the first anti-symmetric mechanical mode, leading to the formation of an optical wavepacket whose comb frequency spacing is locked to the natural frequency of this mode. Figure \ref{fig6} presents the resonator optomechanical responses when the cavity is localized at this position, with an optical loss coefficient of $\kappa=22$ MHz and a pump intensity of $|s_{in}|^2=40$ GHz. The laser detuning is assumed $\omega_l=\omega_c+\omega_2$. Figures 9(b)–9(i) present the system\textquotesingle s two-dimensional sub-phase spaces as obtained from numerical simulations. With the exception of the second mechanical mode, which directly interacts with the optical resonator, thereby establishing an optomechanical energy pathway and yielding a periodic orbit, the displacement-velocity phase planes of the remaining first seven mechanical modes display quasi-periodicity.

\begin{figure}[t!]
	\centering
	\begin{subfigure}[b]{0.94\textwidth}
		\centering
		\includegraphics[width=\textwidth]{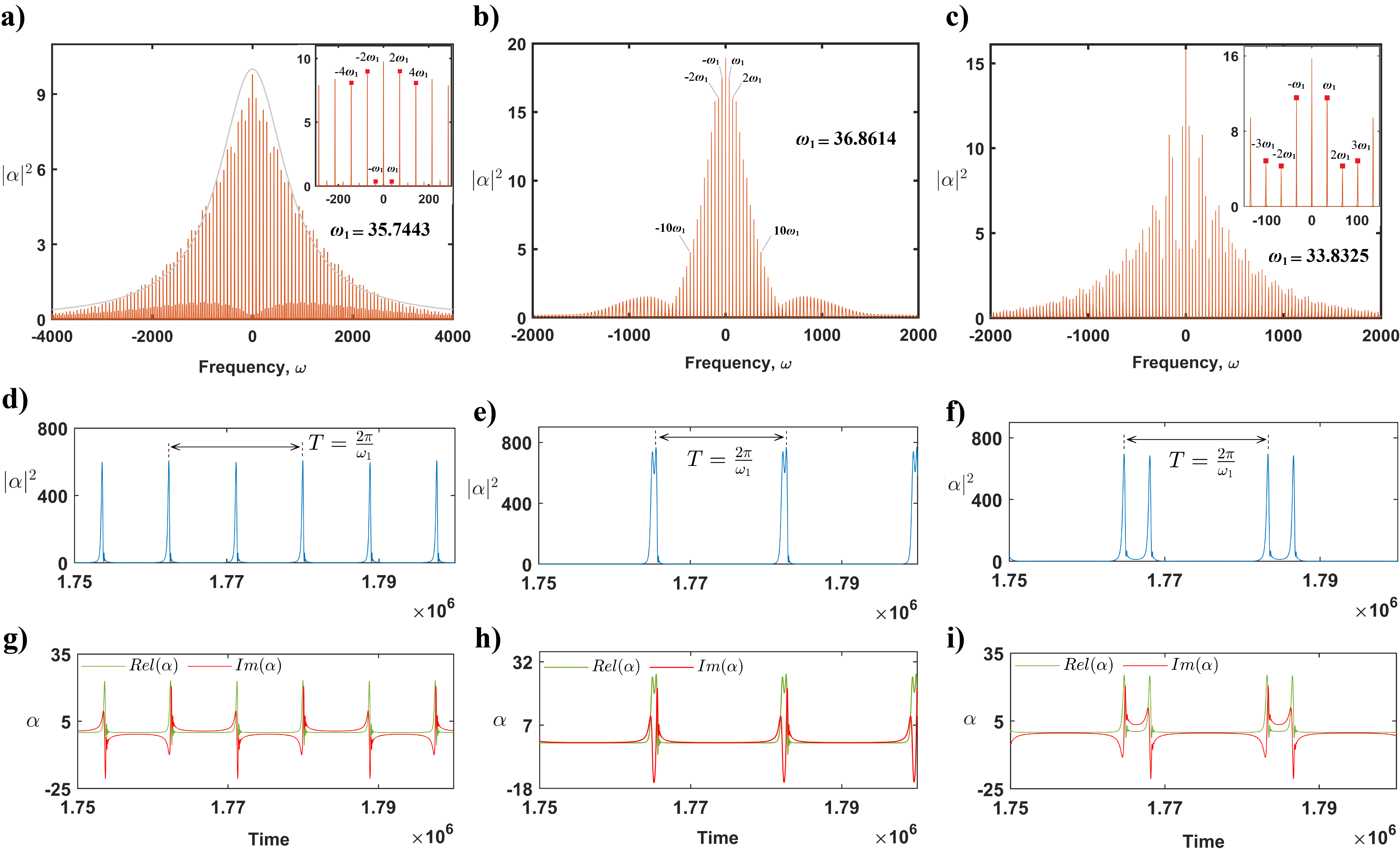}
		\vspace{0.1mm}
	\end{subfigure}
	\caption{Optical response of a cavity localized at the microbeam midspan. (a) Response spectrum without opto-thermal dynamics. (b) and (c) Optical spectra including opto-thermal effects with thermo-mechanical force coefficients $\mu=0.001$ and $\mu=-0.001$, respectively. The optical loss coefficient and input photon flux are fixed at $\kappa=10$ MHz and $|s_{in}|^2=4.356$ GHz. (d)–(f) Time-domain responses corresponding to the spectra in (a)–(c). (g)–(i) Real and imaginary components of the optical mode evolution associated with (d)–(f).}
	\label{fig7}
\end{figure}

Figure \ref{fig7} shows a comparative analysis of the frequency combs generated without and with accounting for opto-thermal dynamics in the optomechanical resonator\textquotesingle s response, for a cavity positioned at the midpoint of the movable mirror. Here, the optical loss coefficient and input photon flux are assumed $\kappa=10$ MHz and $|s_{in}|^2=4.356$ GHz, respectively. Figures \ref{fig7}(a)-\ref{fig7}(c) show the spectrum of the intracavity photon number versus the nondimensional frequency. In the absence of opto-thermal response, the comb power decreases smoothly with distance from the central frequency, yielding a soliton-like spectral profile.
Inspection of the magnitude spectrum reveals that the comb teeth located at even harmonics establish a frequency comb with a spacing of FSR$=2\omega_1$, corresponding to twice the natural frequency of the first bending mechanical mode, Figure \ref{fig7}(a). The mechanical mode\textquotesingle s oscillation frequency is $\omega_1=35.7443$. The inclusion of opto-thermal dynamics introduces slow feedback on the cavity resonance modifying the intracavity field. With a positive thermo-mechanical force coefficient, $\mu=0.001$, the comb bandwidth shrinks, which in turn promotes stabilization of the steady-state response. The comb power associated with odd harmonics becomes comparable to that of even harmonics, resulting in the formation of equidistant spectral lines with a frequency spacing of FSR$=\omega_1$. A positive thermo-mechanical force coefficient signifies that the induced thermal force acts in the positive direction of the mirror's displacement, effectively increasing the microbeam\textquotesingle s curvature and shifting the first bending mode to a higher frequency, $\omega_1=36.8614$, Figure \ref{fig7}(b). Incorporating opto-thermal dynamics with a negative thermo-mechanical force coefficient, $\mu=-0.001$, disrupts the nonlinearity-dispersion balance required for the smooth soliton-like spectral structure, Figure \ref{fig7}(c). With a negative thermo-mechanical coefficient, the induced thermal force acts opposite to the mirror\textquotesingle s positive displacement, leading to a reduction in the microbeam\textquotesingle s curvature and a downward shift of the first bending mode\textquotesingle s natural frequency to $\omega_1=33.8325$. The corresponding optical time histories are presented in Figures \ref{fig7}(d)-\ref{fig7}(f), demonstrating pulse-train responses repetition rates determined by the period of the first mechanical mode natural frequency. The corresponding real and imaginary components of the complex-valued and slowly-varying envelope of the intracavity optical field are displayed in Figures \ref{fig7}(g)-\ref{fig7}(i).

\begin{figure}[b!]
	\centering
	\begin{subfigure}[b]{1\textwidth}
		\centering
		\includegraphics[width=\textwidth]{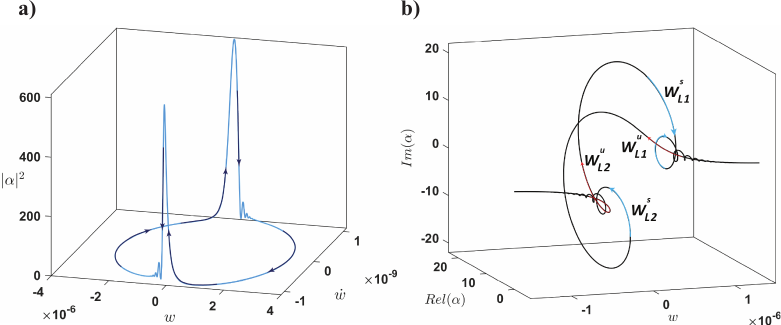}
		\vspace{0.1mm}
	\end{subfigure}
	\caption{Three-dimensional phase spaces corresponding to the results in Figure \ref{fig7}. (a) Optomechanical phase space constructed from the intracavity photon number, microbeam displacement, and velocity at the cavity location ($s_0=0.5$). The system trajectory displays a pulsed response characterized by two sharp peaks, where the width of each peak corresponds to the transition interval between the sub-orbits $L_1$ and $L_2$. (b) System periodic orbit consisting of two sub–Shilnikov-like orbits, $L_1$ and $L_2$, arising in the cavity autonomous dynamics. The stable and unstable manifolds of each sub-orbit are denoted by $W^s$ and $W^u$, respectively.}
	\label{fig8}
\end{figure}

The three-dimensional phase spaces corresponding to the system responses shown in Figures \ref{fig7}(a), \ref{fig7}(d), and \ref{fig7}(g) are constructed using different sets of state variables: Figure \ref{fig8}(a) depicts the microbeam displacement, velocity, and intracavity photon number, while Figure \ref{fig8}(b) illustrates the microbeam displacement together with the real and imaginary components of the optical field amplitude. The system equilibrium is unstable for these values of the input photon flux and optical loss coefficient, and the steady-state response exhibits a stable periodic orbit. The periodic orbit with fundamental periodof $T=2\pi/\omega_1$ consists of two sub-orbits, denoted $L_1$ and $L_2$, each associated with its corresponding invariant manifolds. The stable manifolds, $W^s$, are indicated in blue, while the unstable manifolds, $W^u$, are represented in red: $L_1=\{x\in\mathbb{R}^{2M+2}|x\in W^s\left(L_1\right) \cup W^u(L_1)\}$ and $L_2=\{x\in\mathbb{R}^{2M+2}|\ x\in W^s\left(L_2\right) \cup W^u(L_2)\}$. The system evolves along the stable manifold $W_{L_1}^s$, tracing an inward spiraling trajectory characterized by a pair of complex-conjugate eigenvalues of the linearized system with negative real parts. Along this path, the oscillation amplitude progressively diminishes as the response approaches the vicinity of the point $(a,b,w)\approx(0.015,0.66,{10}^{-6})$.
Once the system state is sufficiently close, this point acts as a repeller, forcing the trajectory to diverge. The response then departs along the unstable manifold $W_{L_1}^u$, a behavior associated with a negative real eigenvalue. The unstable manifold of the first sub-orbit subsequently connects with the stable manifold of the second sub-orbit, where an analogous dynamical scenario unfolds near the point $(a,b,w)\approx(0.015, 0.66,-{10}^{-6})$. Here, $a=Rel(\alpha)$ and $b=Im(\alpha)$. The two sub-orbits generate oscillatory responses reminiscent of those induced by a Shilnikov-type orbit, characterized by a notably brief transition period between successive sub-orbits, thereby producing a sequence of pulse structures in the time domain.

The as-fabricated curved microbeam exhibits commensurate frequency ratios of approximately two-to-one between its first anti-symmetric and first symmetric bending modes ($\omega_2\approx2\omega_1$), as well as between the second symmetric and first anti-symmetric bending modes ($\omega_3\approx2\omega_2$). Consequently, under high input photon flux, the mechanical mode that has the strongest spatial overlap with the optical field, either the first symmetric or the first anti-symmetric bending mode, depending on the cavity localization, can undergo large-amplitude oscillations. The oscillation energy of this dominant mode can then transfer to neighbouring modes via two-to-one nonlinear mechanical energy pathways, thereby exciting them auto-parametrically. The subsequent activation of additional mechanical modes leads to the participation of multiple modes in modulating the cavity resonance, which can drive the optical field toward quasi-periodic dynamics or even chaotic motions (see supplementary videos). Figure \ref{fig9} presents the system dynamics when an optical cavity is localized at $s_0=0.375$, for $\kappa=10$ MHz and $|s_{in}|^2=6.4$ GHz.
\begin{figure}[h]
	\centering
	\begin{subfigure}[b]{0.95\textwidth}
		\centering
		\includegraphics[width=\textwidth]{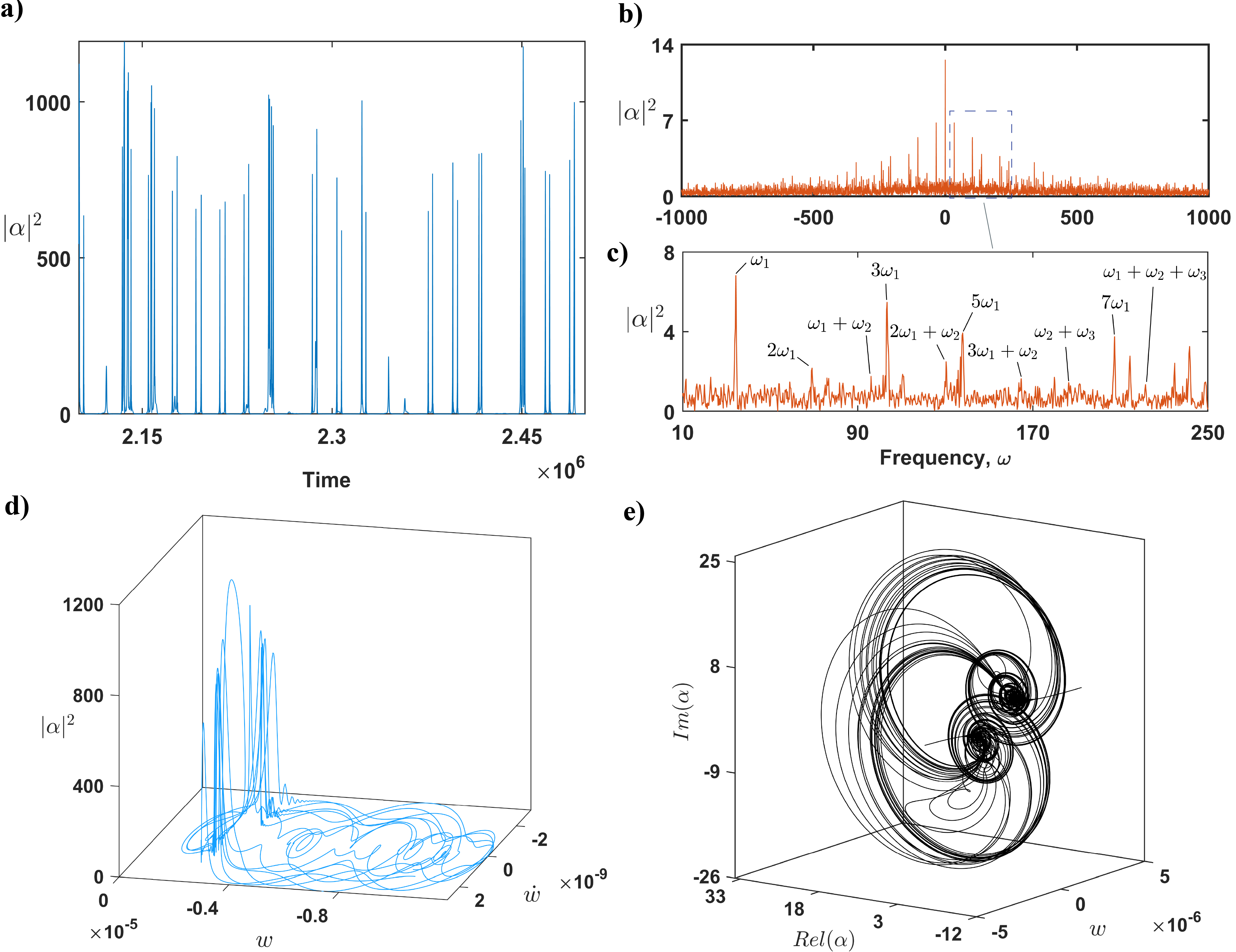}
		\vspace{0.1mm}
	\end{subfigure}
	\caption{Optical mode response of an optomechanical cavity positioned at $s_0=0.375$, for $\kappa=10$ MHz and $|s_{in}|^2=6.4$ GHz. (a) Time-domain evolution of the intracavity photon number. (b) Corresponding spectrum. (c) A magnified view of the spectrum reveals an irregular distribution of frequency components, indicative of chaotic response. Panels (d) and (e) illustrate the corresponding system trajectories in three-dimensional sub-phase spaces.}
	\label{fig9}
\end{figure}
The reduced-order-model is built using the first seven mechanical modes. Integrating the optomechanical equations of motion over a sufficiently long nondimensional time reveals that the intracavity photon number exhibits sharp spikes with irregular temporal spacing, Figure \ref{fig9}(a). The corresponding frequency spectrum displays a noisy profile, with sparsely distributed dominant peaks, associated with the natural frequencies of the mechanical modes that contribute most significantly to variations in the cavity\textquotesingle s optical path length, Figure \ref{fig9}(b). A magnified view of the spectrum within the frequency interval of spanning from 10 to 250, marked by a dashed rectangular, is shown in Figure \ref{fig9}(c). The response contains several dominant frequency components emerging above the noise floor, each corresponding to the natural frequencies of different mechanical modes and to their combinations arising from optomechanical modal interactions.

A selection of representative nonlinear energy exchange pathways is outlined as follows. Once nonlinear energy pathways are established between the optical resonator and the first mechanical mode, mediated by radiation pressure and displacement-induced feedback to the cavity resonance, the intracavity field is initially modulated by this fundamental mode ($|\alpha|^2=A_0+A_1e^{i\omega_1t}+cc$, $q_1=Q_{10}+Q_{11}e^{i\omega_1t}+cc$). Through an energy channel represented by the cubic term, $|\alpha|^2q_1$, present in the radiation pressure force, the oscillation energy of the first mode is subsequently transmitted to the second mechanical mode, leading to the direct excitation of the higher-order mode ($q_2=Q_{20}+Q_{21}e^{i(\omega_2\approx2\omega_1)t}$+cc). The structural quadratic nonlinearity constructed by the first mode, $q_1^2$, in the second mode resonator is effectively negligible, owing to the vanishing projection of the symmetric mode onto the anti-symmetric mode. However, the initial back-action arising from the quadratic mechanical nonlinearity, $q_1q_2$, in the dynamics of the second mechanical mode induces a direct excitation of this mode. Further, the feedback of energy from the first two mechanical modes into the optical resonator, mediated by the cubic optomechanical coupling pathway, $A_0q_1^2q_2$, present in the optical field dynamics, gives rise to a direct excitation of the cavity mode at the combined frequency $2\omega_1+\omega_2$. The frequency-mixing process proceeds through successive optomechanical nonlinear back-actions, whereby higher-order mechanical modes become coupled to the optical resonator, thereby generating additional dominant spectral components. The nonlinear energy exchange between the optical mode and multiple mechanical modes within the autonomous dynamics destabilizes the frequency-locking of the mechanically modulated optical oscillation, potentially driving the resonator into a chaotic regime. The resulting strange attractors of the resonator are illustrated in the three-dimensional phase spaces shown in Figures \ref{fig9}(d) and \ref{fig9}(e). The impact of thermal noise excitation, represented as Gaussian white noise, on the stabilization of the system response merits careful investigation.





\bibliography{sn-bibliography}